\documentclass[prd,aps,onecolumn,preprintnumbers,amsmath,amssymb,superscriptaddress]{revtex4}

\pdfoutput=1

\usepackage{amsmath}
\usepackage{graphicx}
\usepackage{subfigure}
\usepackage{dcolumn}
\usepackage{bm}
\usepackage{amssymb}
\usepackage{latexsym}
\usepackage{epstopdf}
\usepackage{threeparttable}
\usepackage{booktabs}
\usepackage{tabularx}
\usepackage{mathrsfs}
\usepackage{hhline}
\usepackage{multirow}
\usepackage{color}
\usepackage[normalem]{ulem}
\usepackage{xcolor}
\usepackage[T1]{fontenc}
\usepackage{tikz}
\usepackage[colorlinks,linkcolor=magenta,anchorcolor=blue,citecolor=blue]{hyperref}
\def\be{\begin{equation}}
\def\ee{\end{equation}}
\def\bea{\begin{eqnarray}}
\def\eea{\end{eqnarray}}

\begin{document}
\newcommand\redsout{\bgroup\markoverwith{\textcolor{red}{\rule[0.5ex]{2pt}{0.5pt}}}\ULon}




\title{Probing signatures of bounce inflation with current observations}

\author{Shulei Ni}
\email{nishulei@mails.ccnu.edu.cn}
\affiliation{Institute of Astrophysics, Central China Normal University, Wuhan 430079, China}
\affiliation{Key Laboratory of Particle Astrophysics, Institute of High Energy Physics, Chinese Academy of Sciences, Beijing 100086, China}
\author{Hong Li}
\email{hongli@ihep.ac.cn}
\affiliation{Key Laboratory of Particle Astrophysics, Institute of High Energy Physics, Chinese Academy of Sciences, Beijing 100086, China}
\author{Taotao Qiu}
\email{qiutt@mail.ccnu.edu.cn}
\affiliation{Institute of Astrophysics, Central China Normal University, Wuhan 430079, China}
\author{Wei Zheng}
\email{zhengwei@ihep.ac.cn}
\affiliation{Key Laboratory of Particle Astrophysics, Institute of High Energy Physics, Chinese Academy of Sciences, Beijing 100086, China}
\author{Xin Zhang}
\email{zhangxin@mail.neu.edu.cn}
\affiliation{Department of Physics, College of Sciences,
Northeastern University, Shenyang 110004, China}
\affiliation{Center for High Energy Physics, Peking University, Beijing 100080, China}

\begin{abstract}
The aim of this paper is to probe the features of the bouncing cosmology with the current observational data. Basing on bounce inflation model, with high derivative term, we propose a general parametrization of primordial power spectrum which includes the typical bouncing parameters, such as bouncing time-scale, and energy scale. By applying Markov Chain Monto Carlo analysis with current data combination of Planck 2015, BAO and JLA, we report the posterior probability distributions of the parameters. We find that, bouncing models can well explain CMB observations, especially the deficit and oscillation on large scale in TT power spectrum.

\end{abstract}

\maketitle

\section{Introduction}
\label{sec:intro}
The question of what happened in the most beginning of our universe has always been the focus of cosmological research, and in recent years, the high precision measurements on the Cosmic Microwave Background (CMB) as well as other observations provided possibilities for the detailed exploration of the early universe. For instance, the Planck full mission temperature and large scale polarization data measure the spectral index of primordial curvature perturbations to be $n_s=0.968\pm0.006~(1\sigma)$~\cite{Ade:2015xua}, which favors a nearly scale-invariant power spectrum with a slightly red tilt. Comparing the theoretical predictions of inflationary models with the observational constraints on the primordial power spectra, it is found that many inflation models still survive today, but some popular inflation models have been explicitly excluded by the current observations; for the details see the reports of Planck~\cite{Ade:2015xua} and BICEP/Keck~\cite{Array:2015xqh} (see also~\cite{Huang:2015cke,Zhang:2017epd,Guo:2017qjt}), and for relevant theoretical studies see e.g.~\cite{Ijjas:2013vea}.


It is also noted that, at large scales of CMB temperature power spectrum measured by the Planck satellite mission (see the Planck 2015 results~\cite{Ade:2015xua}), there are deficits at both $\ell\lesssim 10$ and $\ell\sim30$, which is already mentioned in WMAP~\cite{Eriksen:2007pc, Hoftuft:2009rq} and Planck 2013~\cite{Ade:2013zuv,Ade:2013nlj}. Moreover, the data points show obvious oscillation trend, although not statistically significant because of the large cosmic variance~\cite{Ade:2013nlj}. It is difficult to interpret these phenomena within the standard framework of ``slow-roll inflation'', indicating that there might be new physics at the early stage of the universe, probably even before inflation. There are several early universe scenarios that can be either supplements or alternatives of the inflation scenario, e.g., the Pre-Big-Bang (PBB) scenario~\cite{Gasperini:1992em, Ghosh:1998bm, Hajian:2010cy, 2014arXiv1408.6441G, Gasperini:2016gre}, the matter bounce scenario~\cite{Wands:1998yp, Finelli:2001sr, Cai:2007zv, Cai:2008qw, Brandenberger:2012zb, Odintsov:2014gea}, the ekpyrotic scenario~\cite{Khoury:2001wf, Khoury:2001bz, Lyth:2001pf, Khoury:2001zk, Lehners:2008vx,Hipolito-Ricaldi:2016kqq}, the bounce inflation scenario~\cite{Piao:2003zm, Piao:2003hh, Cai:2008qb, Liu:2013kea, Xia:2014tda, Qiu:2015nha, Wan:2015hya, Zhang:2007bi, Zhang:2009xp, Zhu:2016dkn}, and so on. Although theoretically the motivation of these scenarios is to avoid the notorious Big-Bang Singularity~\cite{hawking, Borde:1993xh, Borde:1996pt, Nojiri:2017ncd}, phenomenologically these scenarios can also give features on large scales, because the primordial fluctuations can be generated in pre-inflationary phase. In this paper, we will focus on the bounce inflation scenario, which is easy to realize/understand in 4D classical Einstein gravity, without resorting to theories of extra-dimensional spacetime such as string/M-theory, or quantum gravity.

It is not a smooth way at all to build healthy bounce/bounce inflation models. In the original idea of (nonsingular) bounce, one has to make the universe contract to some minimum volume ($H<0$) and then expand again ($H>0$), thus a positive time derivative of the Hubble parameter is needed which, according to the Friedmann Equation, violates the Null Energy condition(NEC)~\cite{Cai:2007qw}. The NEC violation will generally cause the notorious ghost instability problem~\cite{Carroll:2003st, Cline:2003gs}, and things didn't get improved until the work of~\cite{Qiu:2011cy} (see also~\cite{Easson:2011zy}) making use of Galileon/Hordeski theory, which contains higher derivative terms but makes the additional degree of freedom non-dynamical~\cite{Nicolis:2008in, Deffayet:2009wt, Nicolis:2009qm, Deffayet:2010qz, Deffayet:2009mn, Deffayet:2011gz, Horndeski:1974wa}. Moreover, it has been found since 1998~\cite{Wands:1998yp} (see also~\cite{Finelli:2001sr} in 2002) that for single field in contracting phase, the scale-invariant spectrum of primordial perturbations required by the observations can only come out when the equation of state (EoS) in contracting phase is zero, however in that case, as proved in~\cite{Kunze:1999xp, Erickson:2003zm, Xue:2010ux, Xue:2011nw}, the universe will suffer from the anisotropy problem, which cannot be solved unless the EoS of contracting universe is larger than unity. In Ref.~\cite{Qiu:2013eoa} and Ref.~\cite{Qiu:2015nha}, the authors discussed two possibilities of reconciling this contradiction by building models with large EoS in contracting phase, and generate scale-invariant power spectrum using curvaton field and a sequencing inflation period, respectively.
Thus, in this paper, our theoretical analysis will be based on a bounce inflation model within the framework of Horndeski theory~\cite{Qiu:2015nha}.

The evolution of primordial perturbations in bounce inflation scenario will be as follows: Initially, the primordial fluctuations are assumed to be generated in the adiabatic vacuum in the contracting phase. Since the Hubble horizon at that time can be large, all the fluctuation modes reside deeply inside the horizon. If the equation of state in contracting phase is larger than $-1/3$, the horizon will shrink in contracting phase, therefore the fluctuations with larger wavelengths will gradually exit the horizon. If the equation of state in contracting phase is larger than $0$, the fluctuations will get blue-shifted, while those with smaller wavelengths remain inside the horizon. At the bouncing phase when the Hubble parameter passes through $0$, the horizon approaches infinity, all the fluctuations will reenter the horizon again.  At the inflationary phase, the fluctuations will exit horizon as in normal inflation scenario, and reenter the horizon after the end of inflation, thus can be observed by us today. Although in inflationary phase the perturbations do not differ much from the normal inflation case, the pre-inflationary evolutions will be imprinted in the perturbations and encoded in the CMB map, and these information will help us study the pre-inflationary era of the universe, and distinguish between different early universe scenarios. See~\cite{Piao:2003zm, Piao:2003hh, Cai:2008qb, Liu:2013kea, Xia:2014tda, Qiu:2015nha, Wan:2015hya, Cai:2016thi, Cai:2017tku} for pioneering works.

The aim of this paper is to try to find the evidence of the bounce inflation scenario with the current observations.  After the theoretical discussion for providing guidance for parametrization, we will perform the data fitting analysis starting from the typical primordial power spectrum, which contains the characteristics of the evolution of bounce inflation. It is worth pointing out that, in view of the spectral structure of the bounce inflation scenario, the conventional scale invariant spectral parameterization method can not give an efficient diagnosis. Unlike the usual scale invariant spectrum, bounce inflation will provide the primordial curvature perturbation spectrum with characteristic structure,  such as an anomalous depression at large scales and oscillatory behavior at the bounce scale. Numerically, the primordial spectrum of curvature perturbations is more complicated than a power law form, however, artificially it can usually be decomposed into an inflationary inherent power law spectrum, and the part related to the evolution of the contracting phase, i.e. a polynomial combination of lots of parameters, which is usually in front of the amplitude. In this paper, by adopting the Planck full temperature map released in 2015, as well as the observations of baryon acoustic oscillations (BAO) and type Ia supernovae (SN), we determine the posterior distributions and the best-fit values of model parameters, and show their correlations. Moreover, with the best fit values of the parameters, we plot the CMB TT power spectrum to see whether our model is consistent with the anomalies indicated in the observational data.

The rest of the paper is organized as follows: In Sec.~\ref{sec2} we analyze the evolution of background and perturbation of bouncing inflation model. In Sec.~\ref{scalar} we dissect the typical characters in primordial spectrum of bounce inflation model. In Sec.~\ref{sec4}, we study the effects of background parameters on primordial spectrum. In Sec.~\ref{sec5} we perform a global fit analysis on the parameters introduced in Sec.~\ref{sec2}, and make some discussions about the results. Conclusion is given in Sec.~\ref{sec6}. The explicit model realization of bounce inflation is introduced in Appendix. \ref{model}, and for a side check of our model, we also roughly analysis the tensor-scalar-ratio $r$ and the nonlinearity parameter $f_{NL}^{equil}$ in Appendix. \ref{sec_tensor} and \ref{sec_non-gaussian}. \footnote{We thank the referee for pointing this to us.}.

\section{Bouncing inflation model}
\label{sec2}
\subsection{Starting Point: Theoretical Construction}
\label{perturbation}\
We start from a general bounce inflation model in Horndeski theory. 
The Lagrangian is as follows:
\be\label{lagrangian}
L=\sum^5_{i=2}L_i
\ee
where
\begin{equation}
\begin{split}
&L_2={\cal K}(\phi,X)~,\\
&L_3=-{\cal G}_3(\phi,X)\Box\phi~,\\
&L_4={\cal G}_4(\phi,X)R+{\cal G}_{4,X}[(\Box\phi)^2-(\nabla_{\mu}\nabla_{\nu})^2]~,\\
&L_5={\cal G}_5(\phi,X){\cal G}_{\mu\nu}\nabla^{\mu}\nabla^{\nu}\phi-\frac{{\cal G}_{5,X}}{6}[(\Box\phi)^3-3(\Box\phi)^2+2(\nabla_{\mu}\nabla_{\nu})^3]~,
\end{split}
\end{equation}
In above action, the $\cal K$ and ${\cal G}_i$ depend on the scalar $\phi$, $\Box\phi\equiv g^{\mu\nu}\nabla_\mu\nabla_\nu\phi$, ${\cal G}_{i,X}\equiv\partial {\cal G}_i/\partial X$, $X\equiv-g^{\mu\nu}\partial_{\mu}\partial_{\nu}\phi/2$ is the kinetic energy, $R$ is the Ricci scalar and ${\cal G}_{\mu\nu}$ is the Einstein tensor.

In the flat FLRW universe, the metric can be given by
\be
ds^2=-dt^2+a(t)^2\delta_{ij}dx^idx^j~,
\ee
and the Friedmann equations and the equation of motion of $\phi$ are given by
\begin{equation}\begin{split}
&{\cal K}(\phi)X+3{\cal T}(\phi)X^2+3{\cal G}_XH\dot{\phi}^3-2{\cal G}_\phi X+{\cal V}(\phi)-3H^2=0~,\\
&{\cal K}(\phi)X+2{\cal T}(\phi)X^2+\frac{3}{2}{\cal G}_XH\dot{\phi}^3-2{\cal G}_\phi X-{\cal G}_X\ddot{\phi}X+\dot{H}=0~,\\
&[{\cal K}(\phi)+6{\cal T}(\phi)X+6{\cal G}_XH\dot{\phi}+6H{\cal G}_{XX}\dot{\phi}-2({\cal G}_\phi+{\cal G}_{X\phi}X)]\ddot{\phi}+3H[{\cal K}(\phi)+2{\cal T}(\phi)X-2({\cal G}_\phi-{\cal G}_{X\phi}X)]\dot{\phi}\\
&+[2{\cal K}_(\phi)+4{\cal T}_\phi(\phi)X+6{\cal G}_\phi(\dot{H}+3H^2)-2{\cal G}_{\phi\phi}]X-{\cal K}_\phi(\phi)-{\cal T}_\phi(\phi)X^2+{\cal V}_\phi(\phi)=0~.
\end{split}\end{equation}
An explicit model based on the above action is given in~\cite{Qiu:2015nha}. In order to remind the readers without occupying the pages of context, we put the detailed analysis is in the Appendix~\ref{model}.


\subsection{ Background: Parametrization of the model}
Although one can realize bounce inflation in concrete models as shown above, in order to grasp the spirit of bounce inflation without being trapped in model,  we parametrize the scale factor in the bounce inflation scenario as follows:
\bea
\label{a3phase}
a(\eta)=\left\{
\begin{array}{l}
a_{\rm con}(\tilde{\eta}_{B-}-\eta)^{\frac{1}{\epsilon_c -1}}~~~{\rm for}~\eta<\eta_{B-}~,\\\\
a_B[1+\frac{\alpha}{2}(\eta-\eta_B)^2]~~~{\rm for}~~\eta_{B-}\leq\eta\leq\eta_{B+}~,\\\\
a_{\rm exp}(\tilde{\eta}_{B+}-\eta)^{\frac{1}{\epsilon_e -1}}~~~{\rm for}~\eta>\eta_{B+}~,
\end{array}\right.
\eea
where $\tilde{\eta}_{B\pm}$ is defined as $\tilde{\eta}_{B-}\equiv\eta_{B-}-[(\epsilon_c-1){\cal H}_{\rm con}]^{-1}$, $\tilde{\eta}_{B+}\equiv\eta_{B+}-[(\epsilon_e-1){\cal H}_{\rm con}]^{-1}$, where $\eta_{B-}$ ($\eta_{B+}$) is the beginning (ending) time of the bouncing phase, $\eta_B$ is the bouncing point, $a_B$ is the scale factor at $\eta_B$, ${\cal H}_{\rm con}$ (${\cal H}_{\rm exp}$) is the conformal Hubble parameter (${\cal H}\equiv aH$) at $\eta_{B-}$ ($\eta_{B+}$), with $H$ representing the energy scale at the moment, and $\epsilon_c$ ($\epsilon_e$) is the slow-roll parameters in contracting (inflationary) phase. The above parameterization contains all the three (contracting/bouncing/expanding) phases of the bouncing inflation scenario, which we will dub as ``{\it three-phase model}''. We require $\epsilon_c$ to be no less than $3$ (equation of state no less than unity) in order to avoid the anisotropy problem. Moreover, $\epsilon_e$ must be close to $0$ as it works in inflationary phase. A sketch plot of this case is drawn in Fig.~\ref{sketch}.
\begin{figure}
 \centering
\includegraphics[scale=0.35]{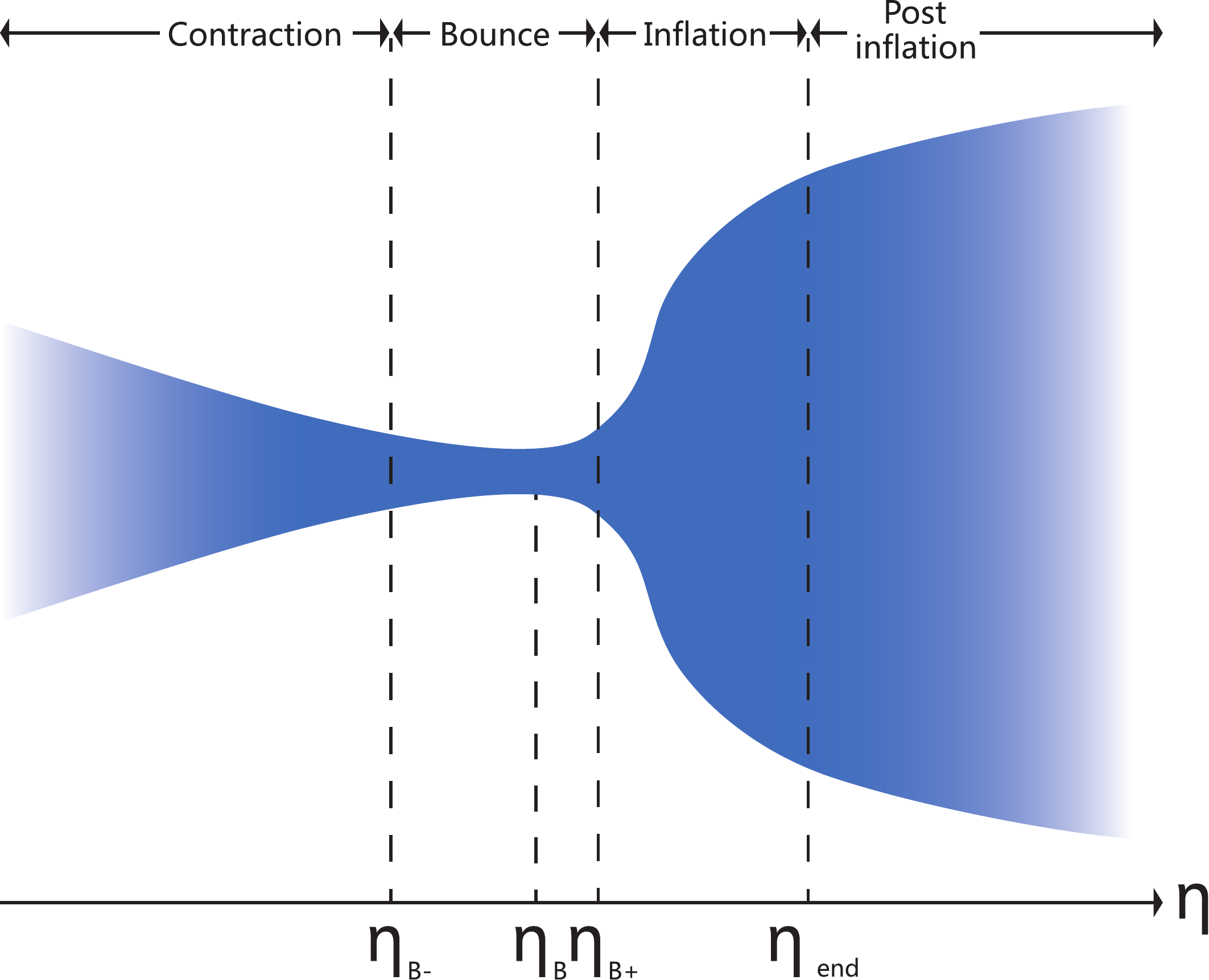}
\includegraphics[width=7cm,height=6cm]{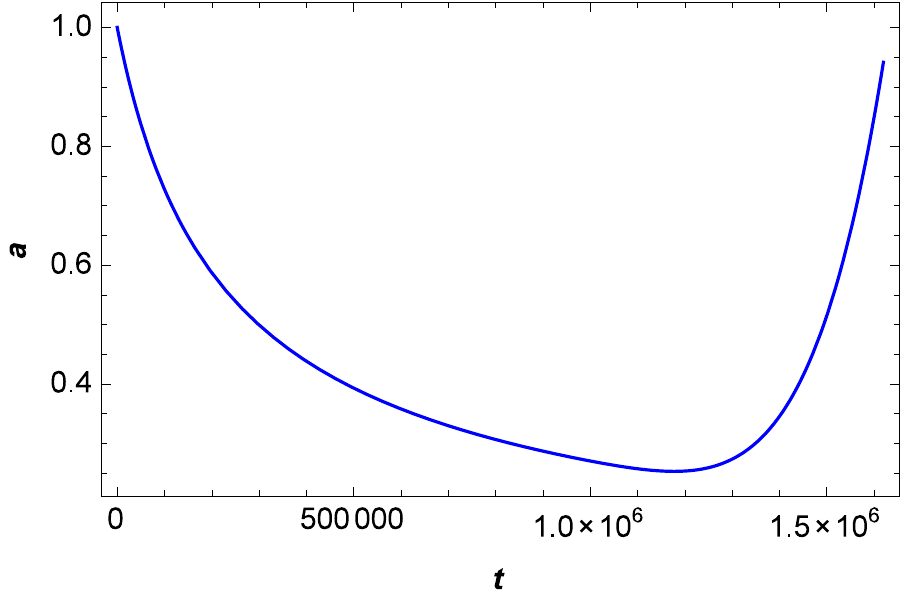}
  \caption{{\it Left:} The sketch plot of the bounce inflation model described by Eq.~(\ref{a3phase}). The universe contracts very slowly at first, and then bounce takes place, after which the universe experiences a fast, inflationary expansion. {\it Right:} The scale factor for the bounce solution with parameter values: $k_0=0.6,\kappa_1=15,t_0=15,\kappa_2=10,\gamma=1\times10^3,\lambda_1=\lambda_2=10,V_0=0.7M_p^4,c=\sqrt{20},\Lambda=1.5\times10^{-2}M_p,v=10M_p$.}\label{sketch}
\end{figure}\\
\section{Cosmological Perturbation}\label{sec3}

\subsection{Scalar Perturbations}\label{scalar}
In order to connect with the observations, in this subsection, we discuss the evolution of perturbation generated in bounce inflation scenario given by Eq.~(\ref{a3phase}) (For similar analysis, see~\cite{Piao:2003zm, Liu:2013kea, Cai:2016thi, Cai:2017tku}). The FLRW metric in an ADM form is:
\be\label{metric_ADM}
ds^2=-N^2dt^2+h_{ij}(dx^i+N^idt)(dx^j+N^jdt)~,
\ee
where $N=1+A$ is the lapse function, $N^i=\partial_i\varphi(i=1,2,3)$ is the shift vector, and $h_{ij}=a^2(t)e^{2\zeta}\delta_{ij}$ is the induced 3-metric. We can rewrite Eq.~(\ref{metric_ADM})
\be\label{metric}
ds^2=-[(1+A)^2-a(t)^{-2}e^{-2\varphi}(\partial\zeta)^2]dt^2+2\partial_i\zeta dtdx^i+a(t)^2e^{2\varphi}d\mathbf{x}^2
\ee
and define the curvature perturbation as
\be
\cal R \equiv \zeta+\frac{H}{\dot{\phi}}\delta\phi~.
\ee

The quadratic action of the curvature perturbation ${\cal R}$ is
\be
S^{(2)}=\frac{1}{2}\int d\eta d^3xa^2\frac{Q}{c^2_s}[{{\cal R}'}^{2}-c^2_s(\partial{\cal R})^2]
\ee
where $'$ is the derivative with respect to conformal time $\eta\equiv\int a^{-1}(t)dt$, and
\begin{equation}\begin{split}\label{QCs}
&Q=\frac{2M_p^4X}{(M_p^2H-{\cal G}_XX\dot{\phi})^2}\left[{\cal K}(\phi)+2{\cal T}(\phi)X+2({\cal G}_X+{\cal G}_{XX}X)\ddot{\phi}+4H{\cal G}_X\dot{\phi}-\frac{2{\cal G}_X^2X^2}{M_p^2}\right]~,\\
&c_s^2=\frac{(M_p^2H-{\cal G}_XX\dot{\phi})^2}{2M_p^4X}\left[{\cal K}(\phi)+6{\cal T}(\phi)X+6H({\cal G}_X+{\cal G}_{XX}X)\dot{\phi}+\frac{6{\cal G}_X^2X^2}{M_p^2}\right]Q~.
\end{split}\end{equation}
Therefore, we can get the equation of motion of the primordial perturbation in contracting phase:
\be
\label{prim_pert}
u''_k+(c_s ^2 k^2 - \frac{z''}{z})u_k = 0~,~~~u_k\equiv z {\cal R}~,~~~z\equiv\frac{a\sqrt{Q}}{c_s}~.
\ee

The scale factor evolves according to Eq.~(\ref{a3phase}) for $\eta<\eta_{B-}$. First of all, we assume that the perturbations are generated in the adiabatic vacuum, which resides deep inside the horizon. The solution is the well-known plane-wave solution:
\be
\label{solvac}
u_k \sim \frac{1}{\sqrt{2k}}e^{-ik\eta}~.
\ee
For simplicity, we assume $Q\simeq2M_p^2\epsilon_c$ and $c_s^2\simeq1$ in contracting phase. As has been shown in the Appendix~\ref{model}, although we need higher derivative term to trigger the bounce, for the regions far away from the bounce, it is possible to make those terms quite suppressed, and the universe behaves like it was driven by a canonical single field. Therefore
\be
\frac{z''}{z} \simeq\frac{a''}{a}\simeq \frac{\mathcal{H}_{\rm{{con}}}^2}{(1+2\mathcal{H}_{\rm{{con}}} (\eta-\eta_{B-}))^2}~.
\ee
Substituting into Eq.~(\ref{prim_pert}) one can get the solution:
\be
\label{sol3c}
u_k=\sqrt{-(\eta-\tilde{\eta}_{B-})}\{c_1H^{(1)}_{\nu_-}[-k(\eta-\tilde{\eta}_{B-})]+c_2H^{(2)} _{\nu_-}[-k(\eta-\tilde{\eta}_{B-})]\}~,~~~\nu_-\equiv\frac{(\epsilon_c-3)}{2(\epsilon_c-1)}~,
\ee
where $H^{(1)} _{\nu_-}$ and $H^{(2)}_{\nu_-}$ are the first and second kind Hankel functions of $\nu_-$ order. Matching Eq.~(\ref{sol3c}) with the vacuum solution Eq.~(\ref{solvac}) one has
\be
c_1=\frac{\sqrt{\pi}}{2}e^{i\frac{\pi}{2}(\nu_-+\frac{1}{2})}~,~c_2=0~,
\ee
and Eq.~(\ref{sol3c}) can be rewritten as:
\be
u_k=\sqrt{-\frac{\pi(\eta-\tilde{\eta}_{B-})}{4}}e^{i\frac{\pi}{2}(\nu_-+\frac{1}{2})}H^{(1)}_{\nu_-}[-k(\eta-\tilde{\eta}_{B-})]~.
\ee
\begin{figure}
\begin{center}
\includegraphics[width=7cm,height=6cm]{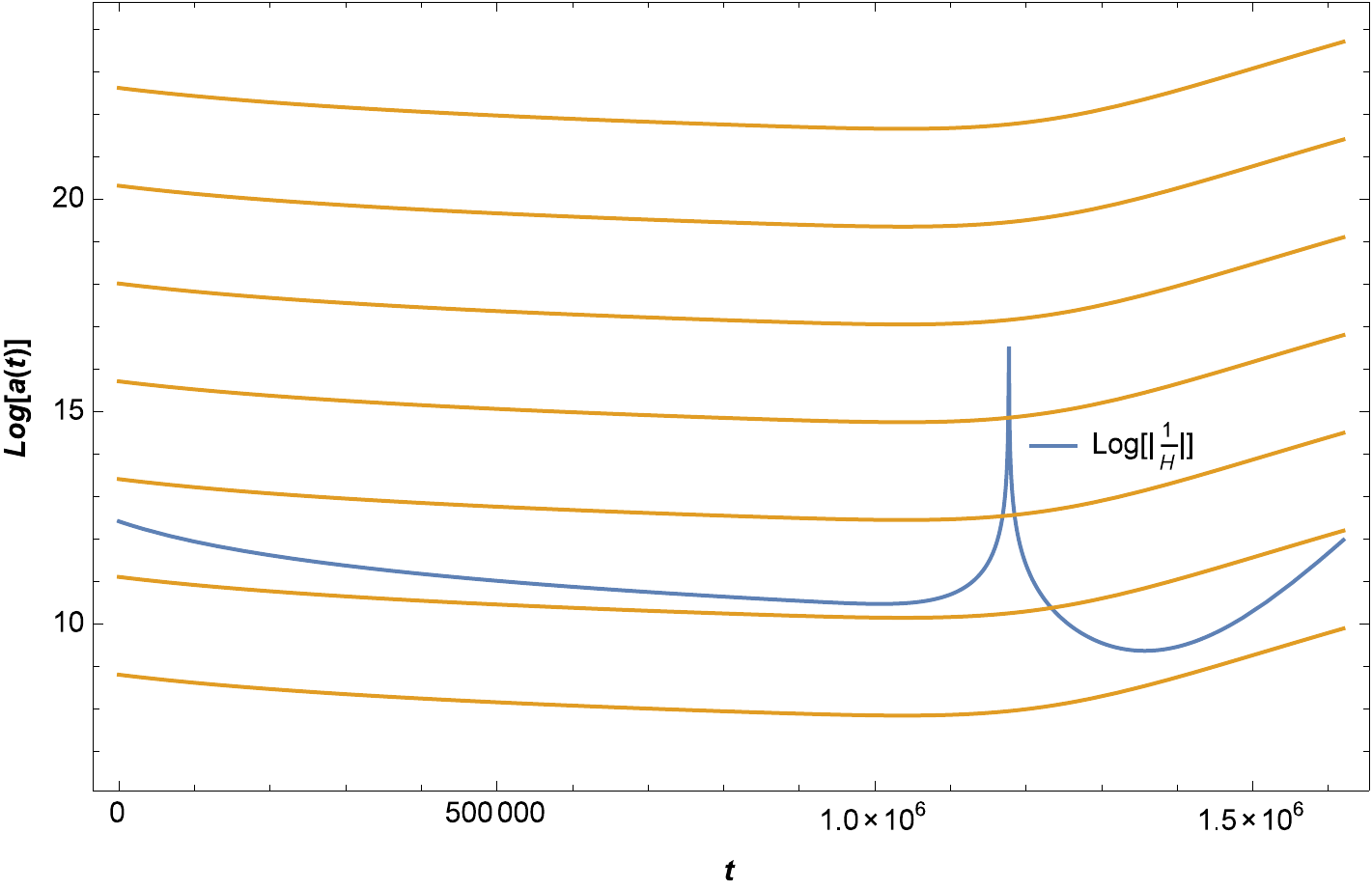}
\includegraphics[width=7cm,height=6cm]{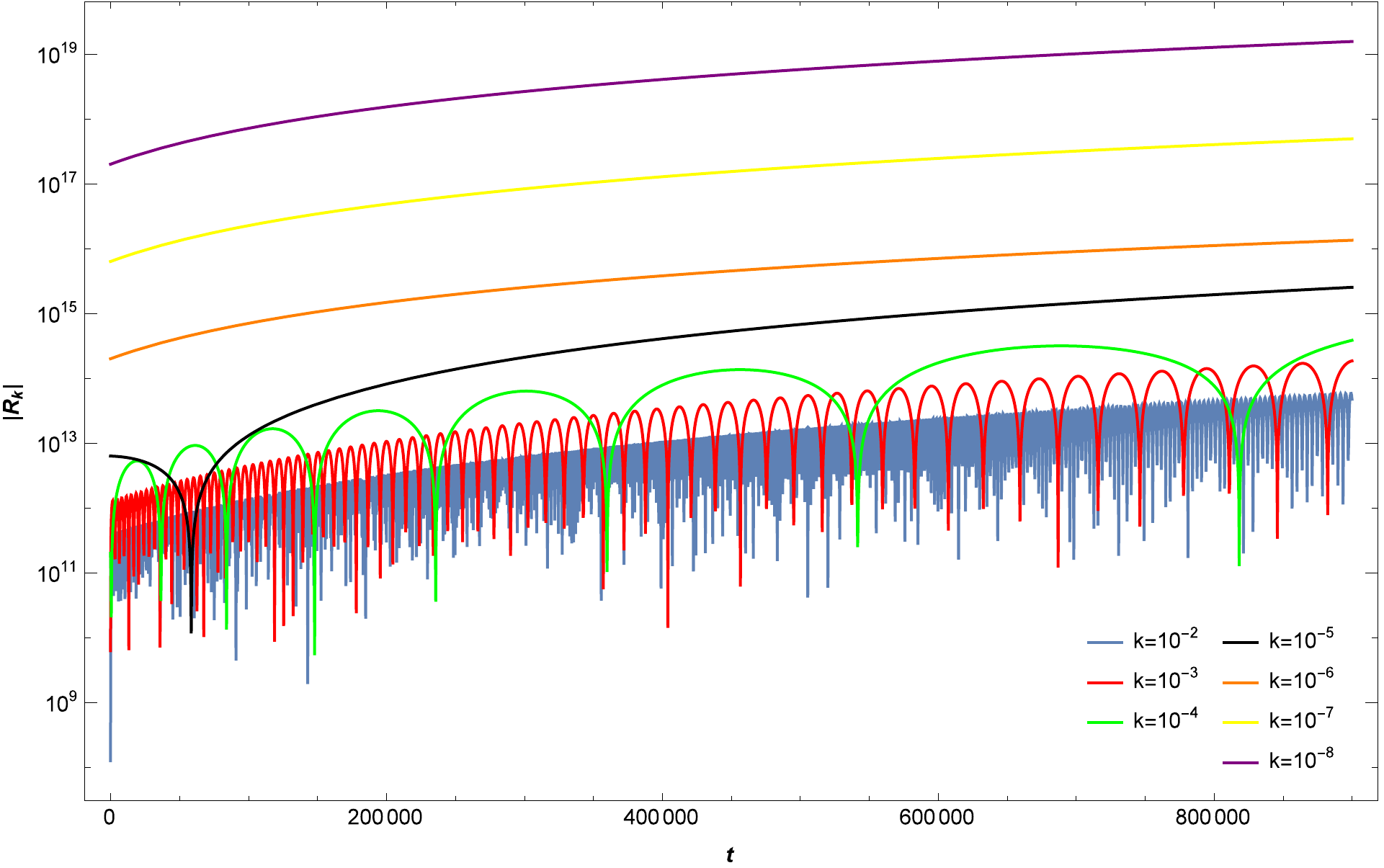}
\caption{ {\it Left:} The evolution of physical Hubble horizon in this model. In the contracting phase, $1/H$ decreases so as the fluctuations with large wavelengths can exit the horizon in this period. Around the bounce time when $H=0$, the horizon diverges and all the modes will reenter into the horizon. In the inflationary phase, the horizon becomes flat since $H$ is close to a constant, so most of the modes will exit the horizon and become classical perturbations. {\it Right:} The evolution of the comoving curvature perturbation for various scales. Large scale modes (small $k$) remain unchanged across the bounce, while at small scale (large $k$), the curvature perturbation has obvious oscillation.}\label{R_k}
\end{center}
\end{figure}

Unlike the contracting phase, during bouncing phase the higher-order derivative terms get involved, making the whole equation of motion much more complicated. Moreover, Null Energy Condition has to be violated to get the bounce. To be specific but without losing generality, we follow~\cite{Qiu:2015nha} to write down the equation of motion of the perturbations in bouncing phase as:
\be
u''_k+[\bar{c}^2_sk^2-(\alpha-\chi)a^2 _B]u_k=0~,
\ee
where $\bar{c}_s$ is effective sound speed during bouncing phase, and $\chi$ is the field-dependent parameter introduced in~\cite{Qiu:2015nha}. The above equation has following solution:
\be
\label{solbounce}
u_k=c_3\cos[l(\eta-\eta_B)]+c_4\sin[l(\eta-\eta_B)]~,
\ee
where $l^2=\bar{c}^2_s k^2 +(\alpha-\chi )a_B ^2$ is a small number, with $\bar{c_s}^2\simeq 0$ and $\alpha\lesssim\chi$. The solution implies that all the perturbation modes are the oscillation modes, i.e., the modes are inside the horizon, which is because during the bouncing phase, $H\rightarrow 0$ and the horizon ($\sim H^{-1}$) approaches to infinity. Moreover, we require that the solution continuously transits from contracting phase to bouncing phase, namely, the solution Eq.~(\ref{solbounce}) should be continuously matched to the solution Eq.~(\ref{sol3c}) at the transition time point $\eta_{B-}$. Making $u_k$ and its derivative continuous at $\eta_{B-}$, the coefficients in the solution in bouncing phase Eq.~(\ref{solbounce}) are:
\begin{equation}\begin{split}
c_3=&\frac{1+i}{4l}\sqrt{\pi}\{\sqrt{-\frac{1}{\mathcal{H}_{\rm{con}}}}kH_1^{(1)}[-\frac{k}{2\mathcal{H}_{\rm{con}}}]\sin[l(\eta_B-\eta_{B-})]\\
&+H_0^{(1)}[-\frac{k}{2\mathcal{H}_{\rm{con}}}]\{\sqrt{-\frac{1}{\mathcal{H}_{\rm{con}}}}l\cos[l(\eta_B-\eta_{B-})]-\sqrt{-\mathcal{H}_{\rm{con}}}\sin[l(\eta_B-\eta_{B-})]\}\}~,\\
c_4=&\frac{1+i}{4l}\sqrt{\pi}\{-\sqrt{-\frac{1}{\mathcal{H}_{\rm{con}}}}kH_1^{(1)}[-\frac{k}{2\mathcal{H}_{\rm{con}}}]\cos[l(\eta_B-\eta_{B-})]\\
&+H_0^{(1)}[-\frac{k}{2\mathcal{H}_{\rm{con}}}]\{\sqrt{-\mathcal{H}_{\rm{con}}}\cos[l(\eta_B-\eta_{B-})]+\sqrt{-\frac{1}{\mathcal{H}_{\rm{con}}}}l\sin[l(\eta_B-\eta_{B-})]\}\}~.\\
\end{split}\end{equation}

After the bounce, the universe will enter into an inflationary expanding phase. The equation of motion of the perturbation is basically the same as Eq.~(\ref{prim_pert}), except that $\epsilon_c$ in $z$ is replaced with $\epsilon_e$, and the scale factor evolves according to Eq.~(\ref{a3phase}) for $\eta>\eta_{B+}$. As in the contracting phase, it is also useful to simplify the perturbation equations by setting $Q\simeq2M_p^2\epsilon_e$ and $c_s^2\simeq1$. Then we have
\be
\frac{z''}{z} \simeq\frac{a''}{a}\simeq \frac{{\mathcal{H}}_{\rm exp}^2}{(1-{\mathcal{H}}_{\rm{exp}}(\eta-\eta_{B+}))^2}~.
\ee
Substituting into Eq.~(\ref{prim_pert}) we get the solution:
\be
\label{solinf}
u_k=\sqrt{-(\eta-\tilde{\eta}_{B+})}\{c_5 H^{(1)}_{\nu_+}[-k(\eta-\tilde{\eta}_{B+})]+c_6H^{(2)}_{\nu_+}[-k(\eta-\tilde{\eta}_{B+})]\}~,~~~\nu_+\equiv\frac{(\epsilon_e-3)}{2(\epsilon_e-1)}~.
\ee
Requiring the continuity of the solution at the transition from bouncing phase to inflationary phase leads to the matching of the solution Eq.~(\ref{solinf}) and Eq.~(\ref{solbounce}) at the transition time point $\eta_{B+}$, giving rise to the explicit expressions of $c_5$ and $c_6$:
\begin{equation}\begin{split}
c_5=&\frac{1}{8k^2}e^{i\frac{\pi}{4}-\frac{ik}{{\mathcal H}_{\rm{exp}}}}\pi\sqrt{-\frac{k}{{\mathcal H}_{\rm{con}}}}\\
&\times\frac{1}{l}({\mathcal H}_{\rm{exp}}^2-i{\mathcal H}_{\rm{exp}}k-k^2)\{kH_1^{(1)}[-\frac{k}{2{\mathcal H}_{\rm{con}}}]+H_0^{(1)}[-\frac{k}{2{\mathcal H}_{\rm{con}}}][l\cos(l\Delta\eta_B)+{\mathcal H}_{\rm{con}}\sin(l\Delta\eta_B)]\}\\
&+({\mathcal H}_{\rm{exp}}+ik)\{-kH_1^{(1)}[-\frac{k}{2{\mathcal H}_{\rm{con}}}]\cos(\Delta\eta_B)+H_0^{(1)}[l\sin(l\Delta\eta_B)-{\mathcal H}_{\rm{con}}\cos(l\Delta\eta_B)]\}~,\\
c_6=&\frac{1}{\sqrt{2}(\frac{k}{{\mathcal H}_{\rm{exp}}})^{3/2}}(\frac{1+i}{8})e^{\frac{ik}{{\mathcal H}_{\rm{exp}}}}({\frac{k}{{\mathcal H}_{\rm{exp}}}})^{3/2}\pi\\
&\times\frac{1}{l}({\mathcal H}_{\rm{exp}}^2-i{\mathcal H}_{\rm{exp}}k-k^2)\{\sqrt{-\frac{1}{{\mathcal H}_{\rm{con}}}}k\sin(l\Delta\eta_B)H_1^{(1)}[-\frac{k}{2{\mathcal H}_{\rm{con}}}]\\
&+H_0^{(1)}[-\frac{k}{2{\mathcal H}_{\rm{con}}}][\sqrt{-\frac{1}{{\mathcal H}_{\rm{con}}}}l\cos(l\Delta\eta_B)-\sqrt{-{\mathcal H}_{\rm{con}}}\sin(l\Delta\eta_B)]\}\\
&+({\mathcal H}_{\rm{con}}-ik)\{-\sqrt{-\frac{1}{{\mathcal H}_{\rm{con}}}}k\cos(l\Delta\eta_B)H_1^{(1)}[-\frac{k}{2{\mathcal H}_{\rm{con}}}]\\
&+H_0^{(1)}[-\frac{k}{2{\mathcal H}_{\rm{con}}}][\sqrt{-{\mathcal H}_{\rm{con}}}\cos(l\Delta\eta_B)+\sqrt{-\frac{1}{{\mathcal H}_{\rm{con}}}}l\sin(l\Delta\eta_B)]\}~.
\end{split}\end{equation}

The numerical evolution of the curvature perturbation $\cal R$ is shown in Fig.~\ref{R_k}. From the left panel we can see that in bounce inflation scenario, the small scale modes will exit the horizon after the bounce, which is same as in standard inflation scenario, while the large scale modes will exit the horizon in contracting phase, which may blueshift the power spectrum. The right panel shows evolution of the curvature perturbation with different wave numbers $k$ in the range $10^{-2}\sim10^{-8}$ with respect to $t$. While the large scale modes $(k=10^{-6},10^{-7},10^{-8})$ behave like a constant, the small scale modes $(k=10^{-2},10^{-3},10^{-4})$ present rapid oscillations~\cite{Battarra:2014tga,Koehn:2015vvy}.

The solution (\ref{solinf}) in the inflationary phase corresponds to the power spectrum that we could observe. The power spectrum of curvature perturbation is defined as:
\be
\label{spectrum}
P_{\cal R}= \frac{k^3}{2\pi^2}|{\cal R}|^2~.
\ee
From the solution Eq.~(\ref{solinf}), we have
\be
\label{spectrum3}
P_{\cal R}=\Delta_{\cal R}^2|c_5-c_6|^2~,
\ee
where
\be
\Delta_{\cal R}^2\equiv\frac{\mathcal{H}^2}{8\pi^2M^2_P \epsilon_e}~,~~~
|c_5-c_6|^2\sim\left\{
\begin{array}{l}
k^{\frac{2\epsilon_c}{\epsilon_c-1}}~,~~~\text{for small $k$}~,\\\\
1+\text{trigonometric functions}~,~~~\text{for large $k$}~,
\end{array}\right.
\ee
where the corrections of trigonometric functions is of sine or cosine type \footnote{For simplicity and illustration, here we chose different matching conditions as in Ref. \cite{Qiu:2015nha}, therefore the result will be somehow different. The similar treatment can be found in e.g. \cite{Piao:2003zm, Liu:2013kea, Cai:2016thi}}. This can cause the oscillation behavior in primordial power spectrum, which comes from the bouncing process, however, it can be averaged to zero for large $k$ and will not effect the amplitude and the spectral index in leading order.

One can parameterize this spectrum (\ref{spectrum3}) as
\be
P_{\cal R}=A_{\rm sIII}(\frac{k}{k_0})^{n_{\rm sIII}-1}~,
\ee
thus the primordial spectrum Eq.~(\ref{spectrum3}) can be described by five free parameters, i.e., ${\cal H}_{\rm con}$, ${\cal H}_{\rm exp}$, $\Delta\eta_B$, $n_{\rm sIII}$, and $A_{\rm sIII}$. The parameters ${\cal H}_{\rm con}$ and ${\cal H}_{\rm exp}$ describe energy of inflation phase and contracting phase, $\Delta\eta_B$ describes the time interval of bouncing process, all the five parameters are necessary.

\section{Effects of background parameters on primordial spectrum}\label{sec4}
In this section, we analyze the typical characters of primordial power spectrum given by the bounce inflation model. We plot primordial power spectrum in Fig.~\ref{Pk} with different background parameters, i.e. ${\cal H}_{\rm con}$, ${\cal H}_{\rm exp}$ and ${\Delta\eta_B}$, and in each plot of Fig.~\ref{Pk}, we only change one bouncing parameter while fix the others in order to highlight the effect from the parameter. Fig.~\ref{Pk-Hcon} gives the effect from ${\cal H}_{\rm con}$, and from that we find ${\cal H}_{\rm con}$ can modulate the spectrum not only on amplitude, but also the locations of wave peaks and troughs, and it is understandable theoretically, since ${\cal H}_{\rm con}$ determines the energy scale of the bounce phase. Fig.~\ref{Pk-eta} shows the effect from $\Delta\eta_B$, which determines the duration of bounce, and we see that once $\Delta\eta_B$ is small, the modulation effect can be neglected. Fig.~\ref{Pk-Hexp} presents the effect come from ${\cal H}_{\rm exp}$ which is the energy scale of the following expansion period, which mainly tilts the amplitude.

Based on the discussions above, we can see that ${\cal H}_{\rm con}$, ${\cal H}_{\rm exp}$ and $\Delta\eta_B$ are the parameters introduced by bouncing process and they characterize the features of primordial power spectrum on large scales. With modulation from those parameters, primordial power spectrum will have a cut off and oscillation in very large scales, which can lead to a depress in CMB TT power spectrum with oscillation, favored by the current Planck data~\cite{Ade:2015xua}. In next section, we will discuss the observational constraints on parameters of the bounce inflation model.

\begin{figure}
\begin{center}
\subfigure[$\hspace{2mm}\Delta\eta=0.0003,{\cal H}_{exp}=7.0$]{
\label{Pk-Hcon}
\includegraphics[scale=0.25]{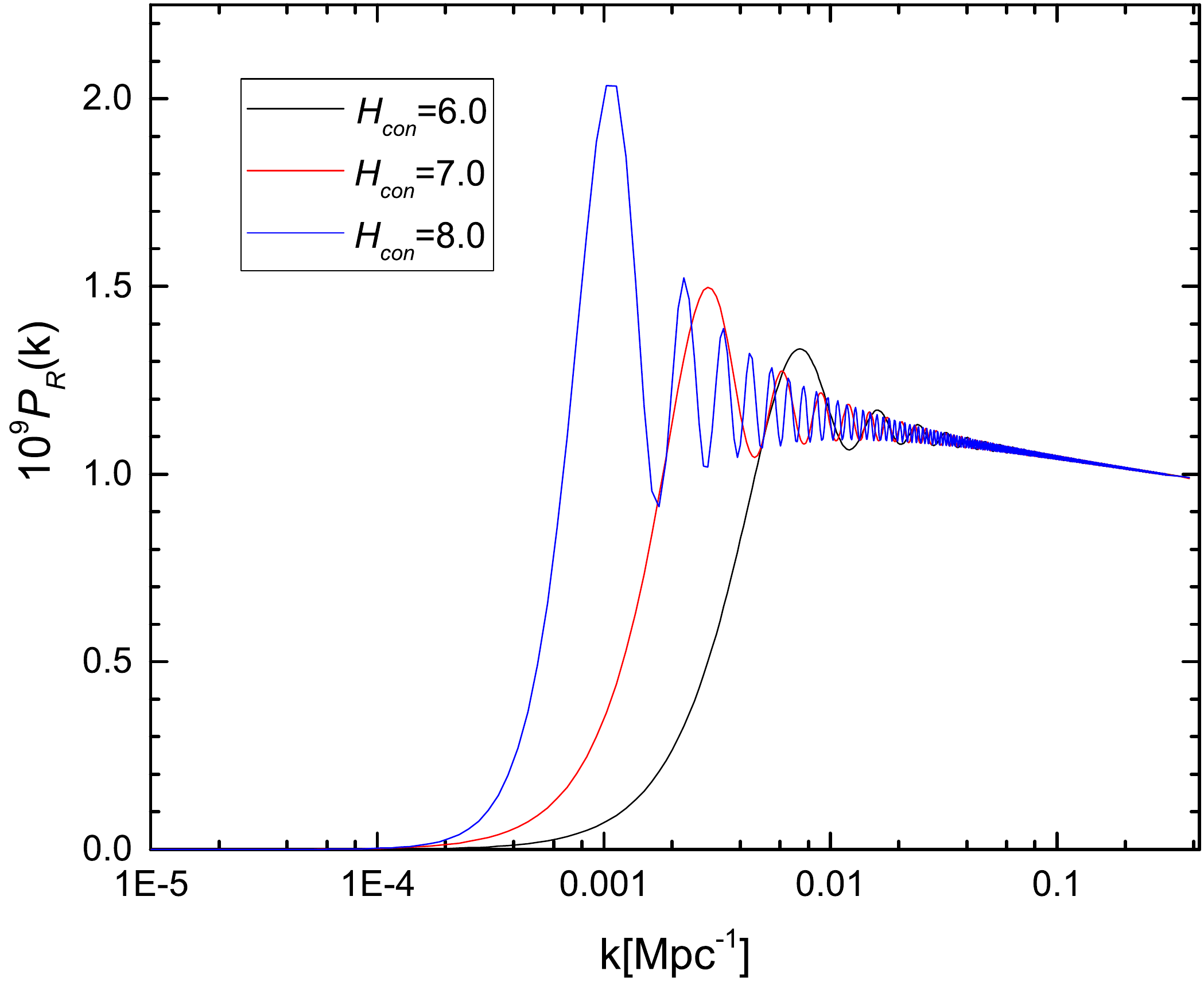}}
\subfigure[$\hspace{2mm}{\cal H}_{con}=7.0,{\cal H}_{exp}=7.0$]{
\label{Pk-eta}
\includegraphics[scale=0.25]{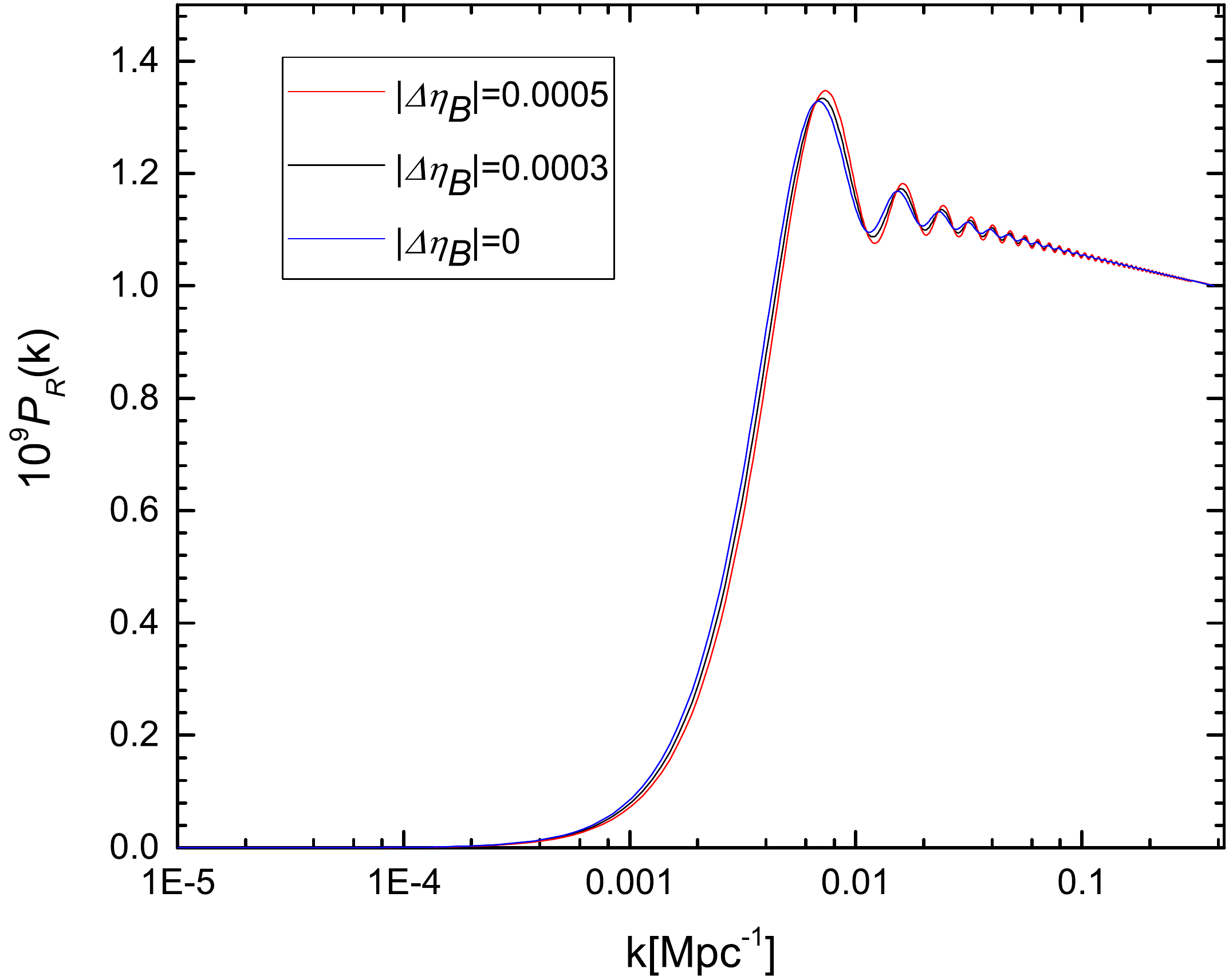}}
\subfigure[$\hspace{2mm}{\cal H}_{con}=7.0,\Delta\eta=0.0003$]{
\label{Pk-Hexp}
\includegraphics[scale=0.25]{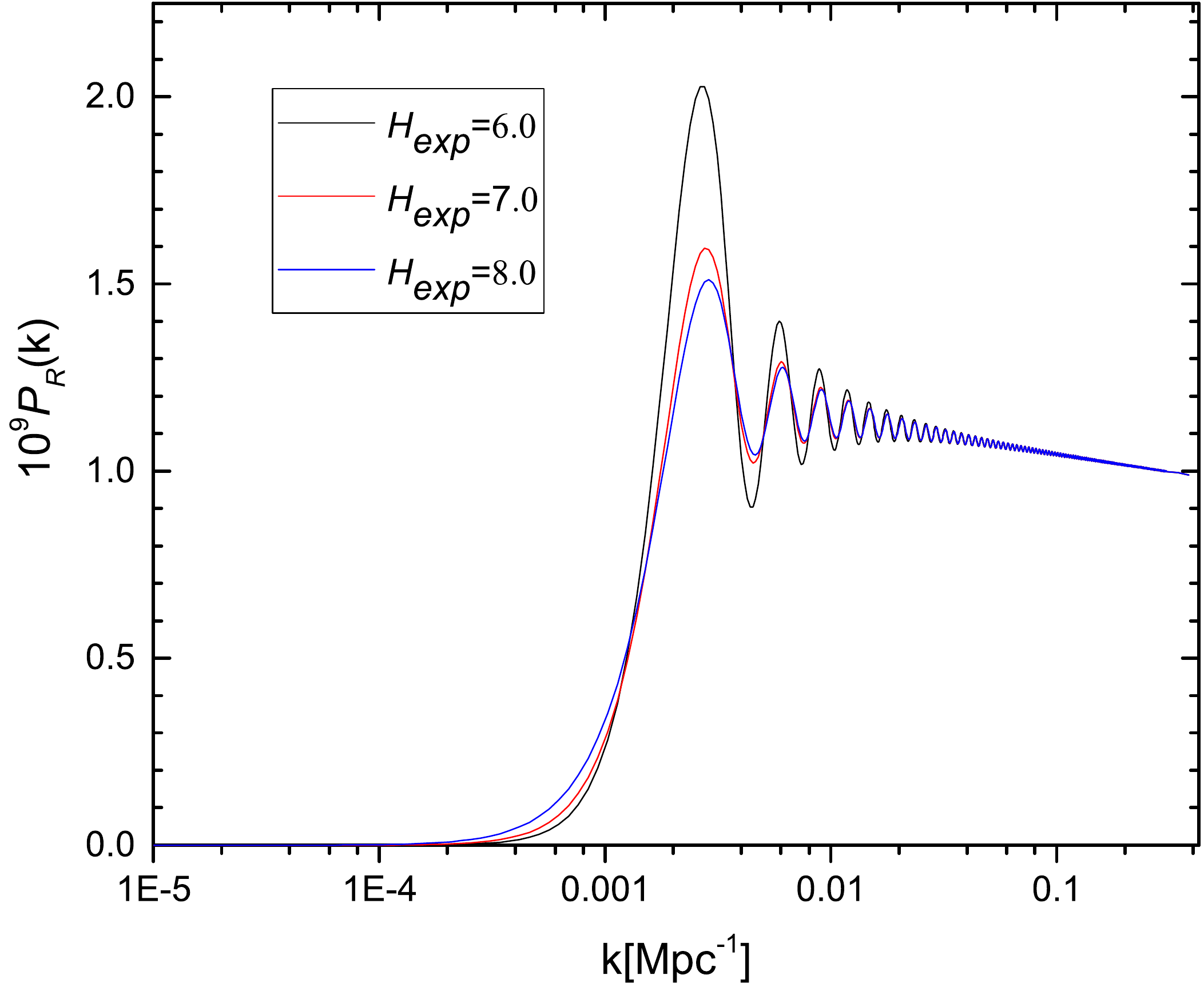}}
\caption{Primordial power spectra of curvature perturbations with different ${\cal H}_{con}$, $\Delta\eta$ and ${\cal H}_{exp}$.}\label{Pk}
\end{center}
\end{figure}

\section{Observational data Diagnosis}\label{sec5}
For observational data diagnosis, we mainly consider CMB temperature power spectrum data~\cite{Ade:2015xua}. CMB is a powerful probe for studying the physics of early universe, which can provide temperature and polarization information about the microwave background photons in the full sky released at the last scattering surface. It measures the angular power spectra of temperature and polarization of CMB photons. We use the Planck 2015 CMB high-$\ell$ ($30\leq\ell\leq2508$) temperature and low-$\ell$ ($2\leq\ell\leq29$) temperature-polarization power spectra data. In order to get better constraints on the background parameters, we adopt BAO data, 6DF, SDSS, WiggleZ and SNIa data of JLA sample in our global fitting~\cite{Cole:2005sx, Betoule:2014frx, Ballardini:2016hpi, Bassett:2009mm, Eisenstein:2005su}.
\begin{figure}
\centering
\includegraphics[scale=0.355]{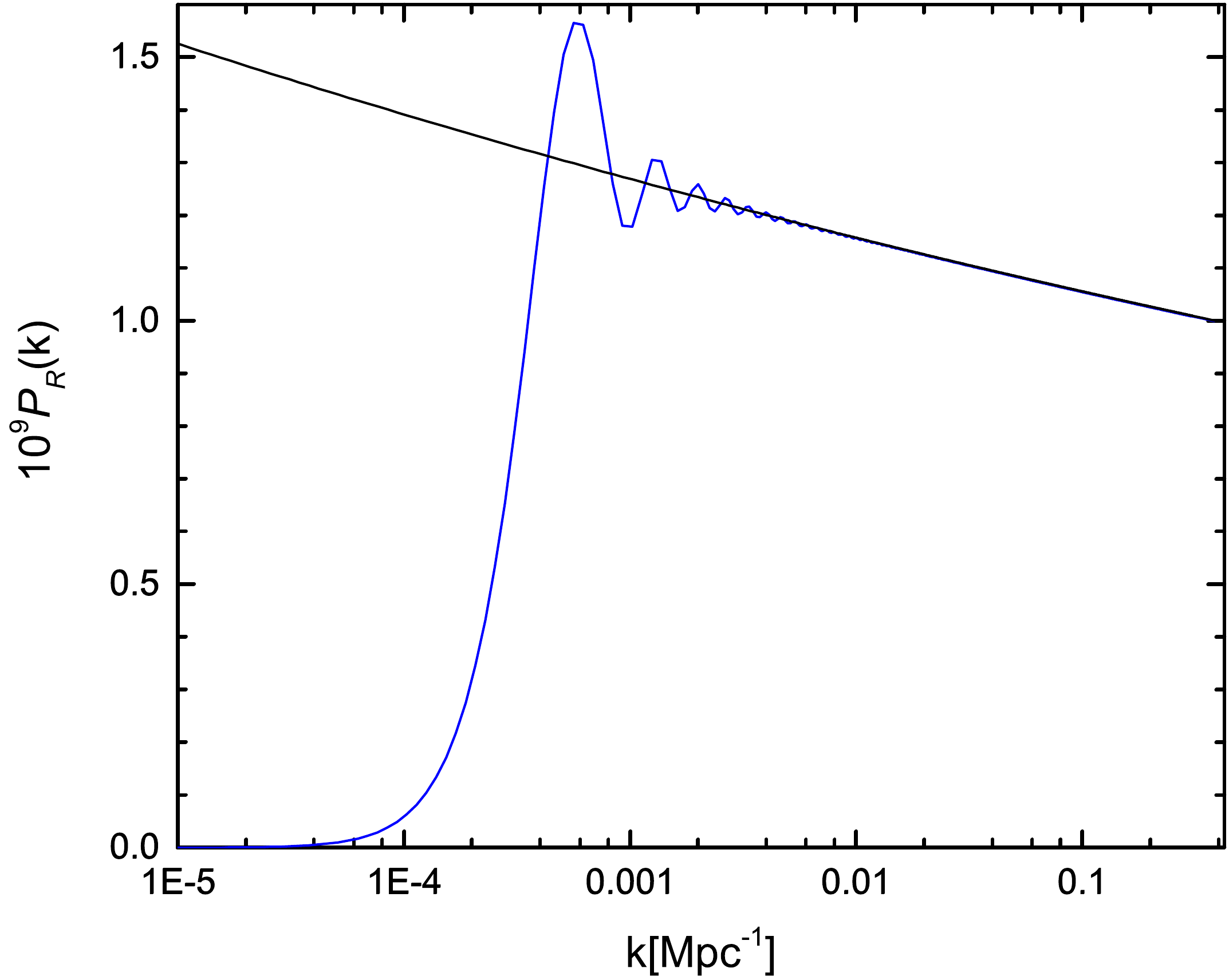}
\includegraphics[scale=0.445]{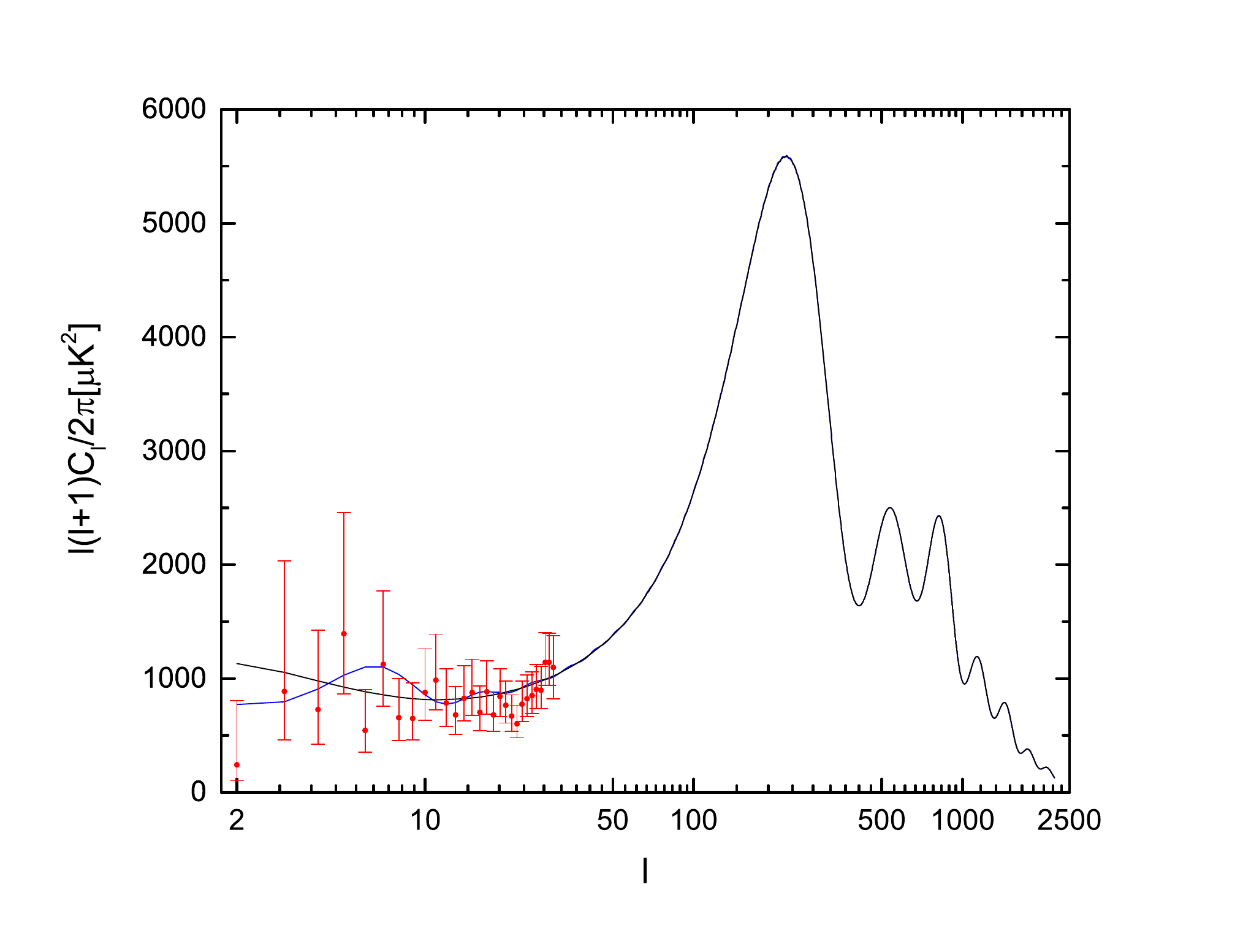}
\caption{Primordial power spectra of curvature perturbations(solid black line) and the CMB temperature angular power spectrum(solid blue line) for the bounce inflation model and the standard inflation model according to the best fits to the current observational data. The red points show the Planck 2015 TT spectrum data with 1$\sigma$ errors.}\label{power}
\end{figure}

We employ the modified CosmoMC program package to perform the global fitting analysis on the bounce inflation model. In combination with basic cosmological parameters of $\Lambda$CDM, the full parameter sets for our model are summarized in Table~\ref{numercial}. In our global fitting analysis, we take ${\cal H}_{\rm con}$, ${\cal H}_{\rm exp}$ as the free parameters, and vary them as other cosmological parameters encoded in MCMC sampling method which adopted by CosmoMC program. We fix $\Delta\eta_B=0$, since it will not affect the primordial power spectrum a lot in a short term bounce phase~\cite{Liu:2013kea}. The model with $\Delta\eta_B=0$ is dubbed as ``two-phase model'' hereafter. The final constraints on primordial power spectrum parameters are listed in Table~\ref{num_B}.

\begin{table}
\centering
\caption{Constraints on the standard inflation model and the three-phase bounce inflation model by using the current Planck+BAO+SN data.}\label{num_A}
\begin{tabular}{l|c|c}
\hline
 $\hspace{2mm}$Parameters$\hspace{2mm}$ &$\hspace{2mm}$Standard model$\hspace{2mm}$&$\hspace{2mm}$Three-phase model$\hspace{2mm}$\\\hline
$\ln{(-{\cal H}_{\rm con})}$&$-$&$<-5.72$\\
$\ln{({\cal H}_{\rm exp})}$&$-$&$<-7.65$\\
$\Delta\eta_B$&$-$&$>-0.0071$\\
$10^{9}A_{\rm sIII}$&$2.176^{+0.099}_{-0.100}$&$2.117^{+0.020}_{-0.011}$\\
$n_{\rm sIII}$&$0.9668^{+0.0185}_{-0.0143}$&$0.9700^{+0.0155}_{-0.0145}$\\ \hline
$\chi^2_{\rm min}$  &11978.60&11973.35\\\hline
\end{tabular}
\end{table}

\begin{table} 
    \centering\footnotesize
    \renewcommand{\tabcolsep}{1pt}
    \begin{threeparttable}

\caption{The block above the middle line shows the basic parameters in the standard $\Lambda$CDM model, and the block below the line includes the derived parameters in the three-phase model and the two-phase model.}\label{numercial} 
\begin{tabular}{llc}
\toprule
\hline
Parameter&Description&Prior range\\
\hline
$\Omega_b h^2$      &      physical baryon density today     &     [0.005,0.1]\\
$\Omega_{dm}h^2$    & physical dark matter density today   &     [0.01,0.99]\\
$\Theta$            & 100 times angular size of sound horizon&     [0.5,10]\\
$\tau$              &      re-ionization optical depth      &     [0.01,0.8]\\
\toprule
\hline
Three-phase model/two-phase model(no $\Delta\eta_B$)\\
\hline
$\ln(-{\cal H}_{\rm con})$           &conformal Hubble parameter at the end of contracting phase&[-12,-7]\\
$\ln({\cal H}_{\rm exp})$           &conformal Hubble parameter at the onset of inflation phase&[-12,-7]\\
$\Delta\eta_B$                   &conformal time length of the bouncing phase&[-0.2,0]\\
$n_{\rm sIII}$               &  scalar spectral index at $k_{\rm{s0}}=0.05{\rm Mpc}^{-1}$& [0.8,1.2]\\
$\ln(10^{10}A_{\rm sIII})$    & amplitude of the primordial curvature perturbations&[2.7,4.0]\\
& at $k_{\rm{s0}}=0.05{\rm Mpc}^{-1}$ &\\
\hline
\toprule
\bottomrule
\end{tabular}
\begin{tablenotes}[para,flushleft]
     \centering
\end{tablenotes}
\end{threeparttable}
\end{table}

We get constraints as $A_s=2.113^{+0.012}_{-0.022}$, $n_s=0.9676^{+0.0154}_{-0.0155}$, ${\cal H}_{\rm con}<-7.00(2\sigma)$ and ${\cal H}_{exp}<-7.51(2\sigma)$. The best fit values of bounce parameters ${\cal H}_{\rm con}$ and ${\cal H}   _{\rm exp}$ are ${\cal H}_{\rm con}=-7.00$ and ${\cal H}_{\rm exp}=-7.51$. We find that the best fits of $\ln(-{\cal H}_{\rm con})$ and $\ln({\cal H}_{\rm exp})$ get very similar values in this case, which indicates that a symmetric bounce process is favored. Theoretically, a symmetric bounce inflation model can be easily achieved, as shown in~\cite{Liu:2013kea}. For comparison, we also constrain the standard inflation model with the primordial power spectrum of the power-law form by using the same observational data, and the fitting results are also shown in Table~\ref{num_A}. Comparing the two models, we find that the bounce inflation model can fit the data better. since we get much smaller $\chi^2$ than the normal $\Lambda$CDM $\Delta\chi^2=5.25$, based on AIC of Bayesian statics for two additional parameters involving in the fitting, which leads to $\Delta\chi^2/\Delta dof=2.625$ larger than 1, meaning that two additional parameters do not lead to overfit. Obviously, introducing bounce parameters is worthy to be paid statistically. For the constraints of the parameters $A_s$ and $n_s$, the two cases are very similar.

With the best fit values given in Table~\ref{num_B}, we plot the primordial power spectrum of curvature perturbations for the bounce inflation model in Fig.~\ref{power}. From the plot we can see that, at large scales (smallest $k$) the spectrum of bounce inflation has an obvious suppression, with the cutoff scale $k\sim0.0005~ {\rm{Mpc^{-1}}}$~\cite{Contaldi:2003zv, Planck:2013jfk, Ade:2015lrj}.  Such a suppression will eventually lead to deficit of the angular power spectrum of temperature at large scales. This is because fluctuations with large scale wavelengths will exit horizon in contracting phase, and will get blue-tilted from the solution (~\ref{sol3c}). After the primordial perturbation evolves to the bounce scenario, the primordial power spectrum has a damped oscillation at $0.0005 ~{\rm{Mpc^{-1}}}<k<0.005 ~{\rm{Mpc^{-1}}}$~\cite{Jackson:2013vka,Meerburg:2013dla,Planck:2013jfk,Ade:2015lrj}, which is due to the fact that all the fluctuation modes will reenter the horizon around the bounce point. The oscillations within the bouncing scenario might explain the anomalous behavior of the CMB spectrum at $20<l<40$. When $k>0.005~{\rm{Mpc^{-1}}}$, the universe bounces into standard inflation phase that has a nearly scale-invariant power spectrum with a slightly red tilt.

We also plot the CMB temperature angular power spectrum for the bounce model comparing with the standard inflation model according to their best-fit values, and the observational data at all scales. We see that the bounce inflation model can realize suppression and oscillations of $C_\ell$ spectrum at large scales, and can fits the data well with the oscillating modulation. We also try to free $\Delta\eta_B$ in global fitting analysis, and we get an upper limit on $\Delta\eta_B$, the best fit value is consist with 0, indicating that the bouncing process duration can be very short. If the bouncing phase is short enough, namely $|\Delta\eta_B|\rightarrow0$~\cite{Liu:2013kea}, as shown in Fig.~\ref{three-stage_1D}. We also plot the 2D posterior distribution contours for the parameters in the $A_{sII}-{\cal H}_{con}, n_{sII}-{\cal H}_{con}, A_{sII}-{\cal H}_{exp}$ and $n_{sII}-{\cal H}_{exp}$ planes in Fig.~\ref{two-stage_2D}. During our calculation, we find freeing $\Delta\eta$ will not decrease $\chi^2$ a lot, i.e. $\Delta\chi^2=\chi^2_3-\chi^2_2=-0.3$, comparing with freeing $\cal{H}_-$ and $\cal{H}_+$. Thus, we need more observational data for tighter constraints.

\begin{figure}
\centering
\includegraphics[width=12cm,height=4cm]{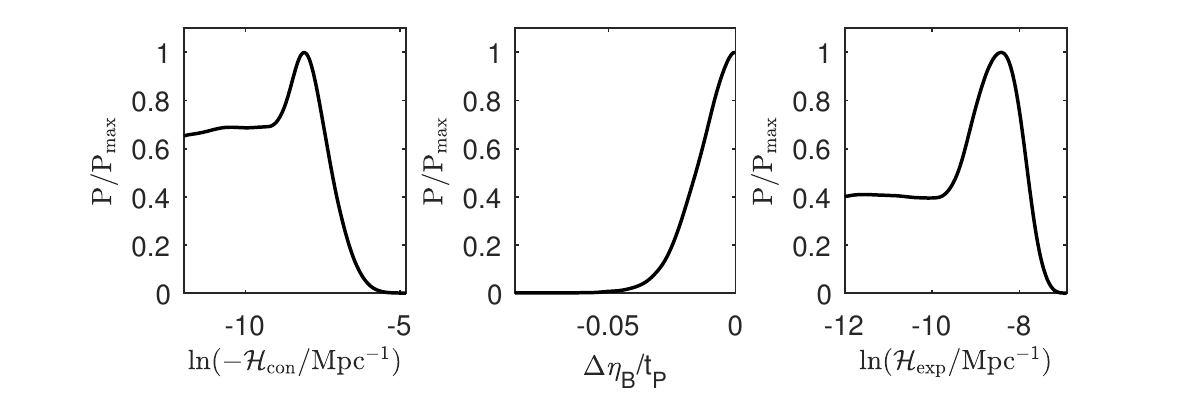}
\caption{One dimensional marginalized posterior distributions for ${\cal H}_{con}, \Delta\eta_B$ and ${\cal H}_{exp}$ from the current Planck+BAO+SN data.}\label{three-stage_1D}
\end{figure}
\begin{figure}
\centering
\includegraphics[scale=0.3]{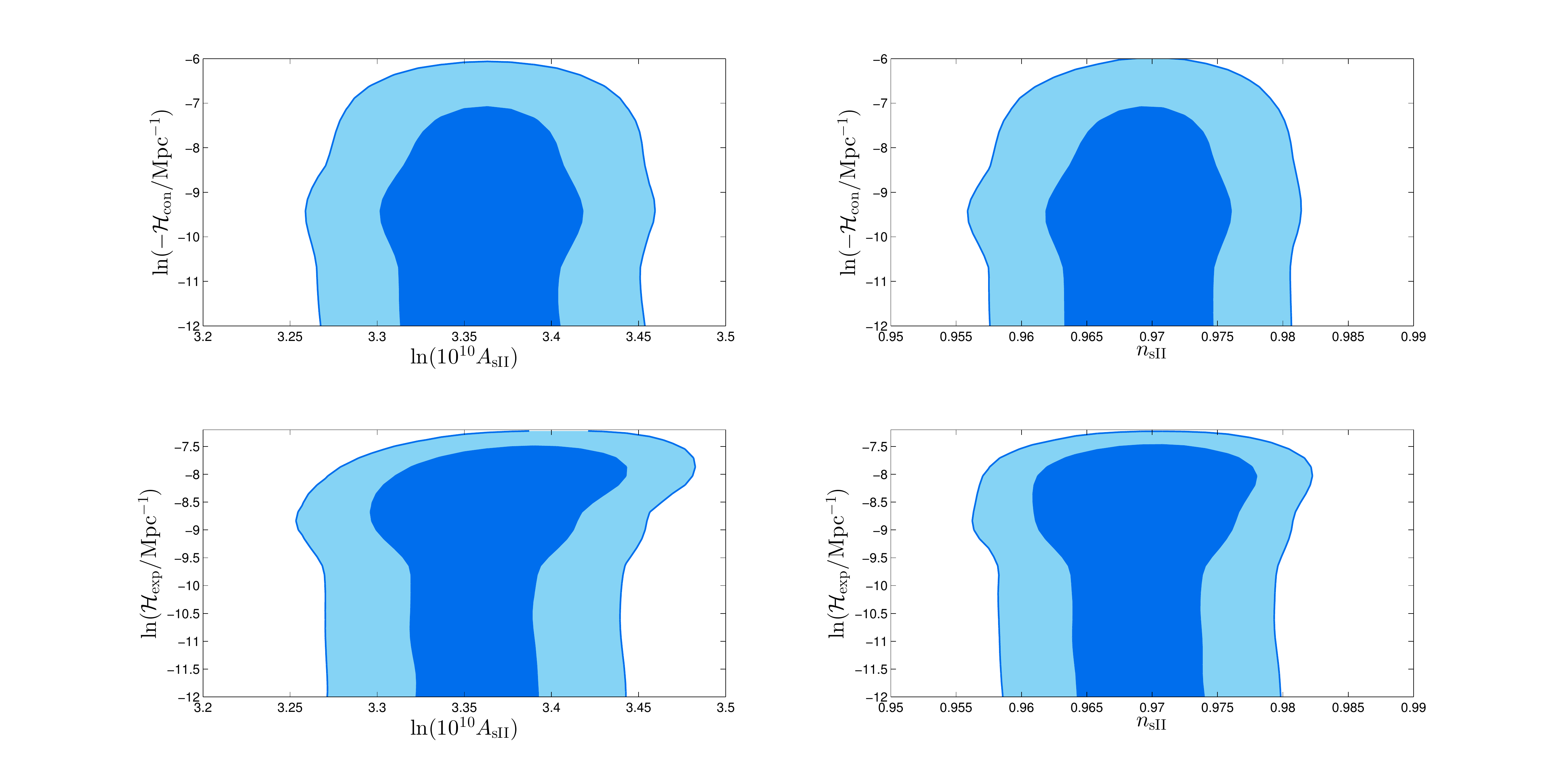}
\caption{Two dimensional joint marginalized constraints (68\% and 95\% confidence levels) on ${\cal H}_{\rm con}$, ${\cal H}_{\rm exp}$, $A_{\rm sII}$ and $n_{\rm sII}$ from the current Planck+BAO+SN data.}\label{two-stage_2D}
\end{figure}



\begin{table}
\centering
\caption{Constraints on the standard inflation model and the two-phase bounce inflation model by using the current Planck+BAO+SN data.}\label{num_B}
\begin{tabular}{l|c|c}\hline
 $\hspace{2mm}$Parameters$\hspace{2mm}$ &$\hspace{2mm}$Standard model$\hspace{2mm}$&$\hspace{2mm}$Two-phase model$\hspace{2mm}$\\\hline
$\ln{(-{\cal H}_{\rm con})}$&$-$&$<-7.00$\\
$\ln{({\cal H}_{\rm exp})}$&$-$&$<-7.51$\\
$10^{9}A_{\rm sII}$&$2.176^{+0.099}_{-0.100}$&$2.113^{+0.012}_{-0.022}$\\
$n_{\rm sII}$&$0.9668^{+0.0185}_{-0.0143}$&$0.9676^{+0.0154}_{-0.0155}$\\ \hline
$\chi^2_{\rm min}$  &11978.60&11973.65\\\hline
\end{tabular}
\end{table}

\section{Conclusion}\label{sec6}
Standard cosmological scenario of Big-Bang and inflation has achieved great success. However, some other scenarios are still not excluded. Especially, to explain the anomalies in CMB observation as well as to solve theoretical problems such as singularity, it is interesting to take the alternative theories into account.

In this paper, we study the scenario in which a bounce happened before inflation. We consider the ``three-phase'' (or ``two-phase'' after setting $\Delta\eta_B=0$) parameterized bounce inflation models, which can be modeled by the Horndeski theory. We derived the primordial power spectra for the model in a rather general form, and showed that the results are determined by the parameters ${\cal H}_{\rm con}, {\cal H}_{\rm exp}, n_s, A_s$ (two-phase model), and an additional parameter, $\Delta\eta_B$ (three-phase model). Using the data combination of Planck 2015, BAO and JLA, we placed the observational constraints on these parameters, and determined their 1D posterior distributions and 2D posterior contours. Using the best fit values, we plotted the primordial power spectrum and the CMB TT spectrum, and showed that the suppression of the spectrum at large scales and the oscillation behavior at mediate scales can well explain the anomalies in the CMB observational data, which is a support for the bounce inflation scenario. We also roughly calculate the tensor-scalar-ratio $r$ and the nonlinearity parameter $f_{NL}^{equil}$. Moreover, we found that the correlation between the comoving Hubble parameter (during the costracting phase or inflation phase) and $n_s$ and $A_s$ are weak.

The results of the ``two-phase model'' and ``three-phase model'' don't differ too much, and the addition of $\Delta\eta_B$ cannot give better constraints for the current observational data, which means that we still cannot make further probe to the bounce process itself, or distinguish between the two models. In order to do so, we need to expect more precise observational data, especially on large scales, in the future.




\begin{acknowledgments}
We thank Hai-guang Li, Jun-Qing Xia, Siyu Li, Yang Liu, Yongping Li, Yun-Song Piao and Ze Luan for useful discussions. The work is supported by the National Natural Science Foundation of China (Grants No.~11405069, No.~11522540, No.~11653001, No.~11653002 and No.~11690021), and the National Program for Support of Top-notch Young Professionals, and the Provincial Department of Education of Liaoning (Grant No.~ L2012087) and the youth innovation promotion association project of the Chinese Academy of Sciences (CAS), the Outstanding young scientists project of the CAS, and the Strategic Priority Research Program of the CAS (Grant No. XDB23020000).

\end{acknowledgments}

\appendix
\section{An Explicit Bounce Inflation Model}
\label{model}
In this section, we review how the parameterization of Eq.~(\ref{a3phase}) can be realized by a realistic model. Ref.~\cite{Qiu:2015nha} provide an interesting example of such a realization, the Lagrangian of which is:
\be
L={\cal K}(\phi)X+{\cal T}(\phi)X^2-{\cal G}(X,\phi)\Box\phi-{\cal V}(\phi)~,
\ee
where $X\equiv -\nabla_\mu\phi\nabla^\mu\phi/2$, $\Box\equiv\nabla_\mu\nabla^\mu$, and we have made use of the mechanism of Galileon theories in order to get rid of the ghost instability problem~\cite{Qiu:2011cy, Easson:2011zy, Nicolis:2008in, Deffayet:2009wt, Nicolis:2009qm, Deffayet:2010qz}. The shape functions ${\cal K}(\phi)$, ${\cal T}(\phi)$, ${\cal G}(X,\phi)$ and the potential ${\cal V}(\phi)$ are chosen to be:
\bea
\label{function}
&&{\cal K}(\phi)=1-\frac{2k_0}{[1+2\kappa_1(\phi/M_p)^2]^2}~,~{\cal T}(\phi)=\frac{1}{M_p^4}\frac{t_0}{[1+2\kappa_2(\phi/M_p)^2]^2}~,~{\cal G}(X,\phi)=\frac{1}{M_p^3}\frac{\gamma X}{[1+2\kappa_2(\phi/M_p)^2]^2}~, \\
\label{potential}
&&{\cal V}(\phi)=[1-\tanh(\lambda_1\frac{\phi}{M_p})]V^{con}(\phi)+[1+\tanh(\lambda_2\frac{\phi}{M_p})]V^{inf}(\phi)~,\nonumber\\
&&V^{con}(\phi)=-V_0 e^{c\phi/M_p}~,V^{inf}(\phi)=\Lambda^4(1-\frac{\phi^2}{v^2})^2~,
\eea
where $k_0$, $\kappa_1$, $t_0$, $\kappa_2$, $\gamma$, $\lambda_1$, $\lambda_2$, $V_0$, $c$, $\Lambda$, and $v$ are constants, and $V^{con}$ and $V^{inf}$ are the part of potential in contracting phase and inflationary phase, respectively. In order to find out the bounce inflation in detail, we plot ${\cal K}(\phi),{\cal T}(\phi),X^{-1}{\cal G}(X,\phi)$, ${\cal V}(\phi)$ and $\phi(t)$ numerically in Fig~\ref{phi} and Fig~\ref{Vphi}. From the Fig~\ref{Vphi} we can see that the shape of the potential shares the same shapes of $V^{con}(\phi)$ for negative $\phi$ while that of symmetry breaking inflation potential $V^{inf}(\phi)$ for positive $\phi$. We will see below that $\phi=0$ is almost the division of contracting and expanding phases. The parameters are chosen as $k_0=0.6$, $\kappa_1=15$, $t_0=5$ $\kappa_2=10$, $\gamma_1=1\times10^{-3}$, $\lambda_1=\lambda_2=10$, $V_0=0.7M_p^4$, $c=\sqrt{20}$, $\Lambda\approx1.5\times10^{-2}$, $10M_p$. By such a choice, at the region far from the bounce (where we set as $|\phi/Mp|\gg 1$), ${\cal K}(\phi)$ goes to unity while ${\cal T}(\phi)$ and ${\cal G}(X,\phi)$ are turned off, and the Lagrangian reduces to that of two-stage canonical single field:
\be\label{limit}
{\cal L}^{con}=X-V^{con}(\phi)~,~~~{\cal L}^{inf}=X-V^{inf}(\phi)~,
\ee
the former of which is just the Lagrangian of the ekpyrotic model with $w_c\geq 1$ ($\epsilon_c\geq 3$), while the latter is just the Lagrangian of the symmetry breaking inflation model with $w_c\geq 1$ ($\epsilon_c\geq 3$), giving rise to the parameterization of Eq.~(\ref{a3phase}) for $\eta<\eta_{B-}$ ($\eta>\eta_{B+}$). On the other hand, at the region near the bounce ($|\phi/M_p|\ll 1$) where the higher-order derivative terms take part in, it is difficult to have reduced Lagrangian and equation of motion. However, from the numerical plot in~\cite{Qiu:2015nha} one can mimic the Hubble parameter with the function $H=\alpha t$ with positive coefficient $\alpha$, which can get the parametrization of Eq.~(\ref{a3phase}) for $\eta_{B-}\leq\eta\leq\eta_{B+}$.
\begin{figure}
\begin{center}
\includegraphics[scale=0.6]{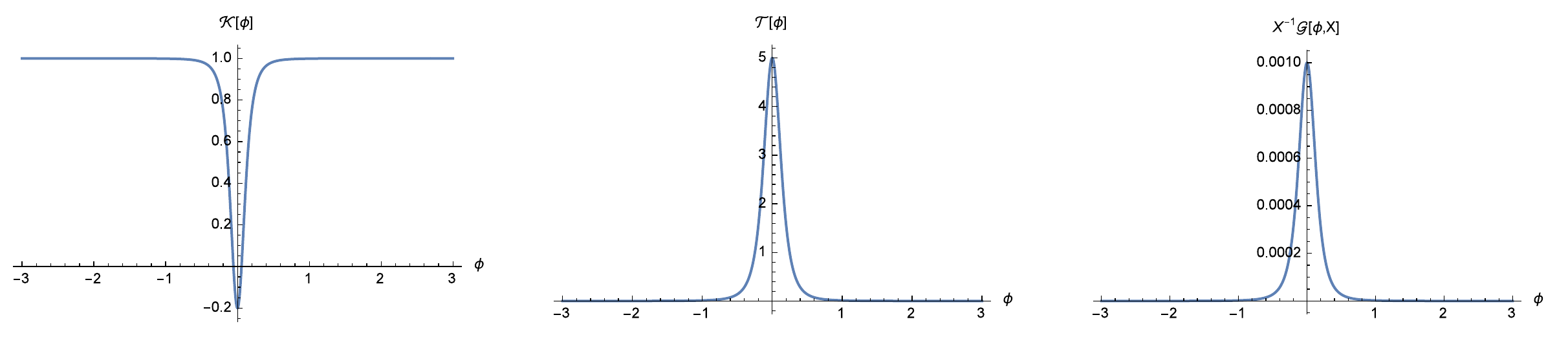}
\caption{Plots of functions ${\cal K}(\phi)$, ${\cal T}(\phi)$ and $X^{-1}{\cal G}(\phi)$ in Eq.~(\ref{potential}). In such a choice, all the three functions have nontrivial value only around the bouncing point, which is useful to trigger the bounce}\label{phi}
\end{center}
\end{figure}

\begin{figure}
\begin{center}
\includegraphics[width=7cm,height=6cm]{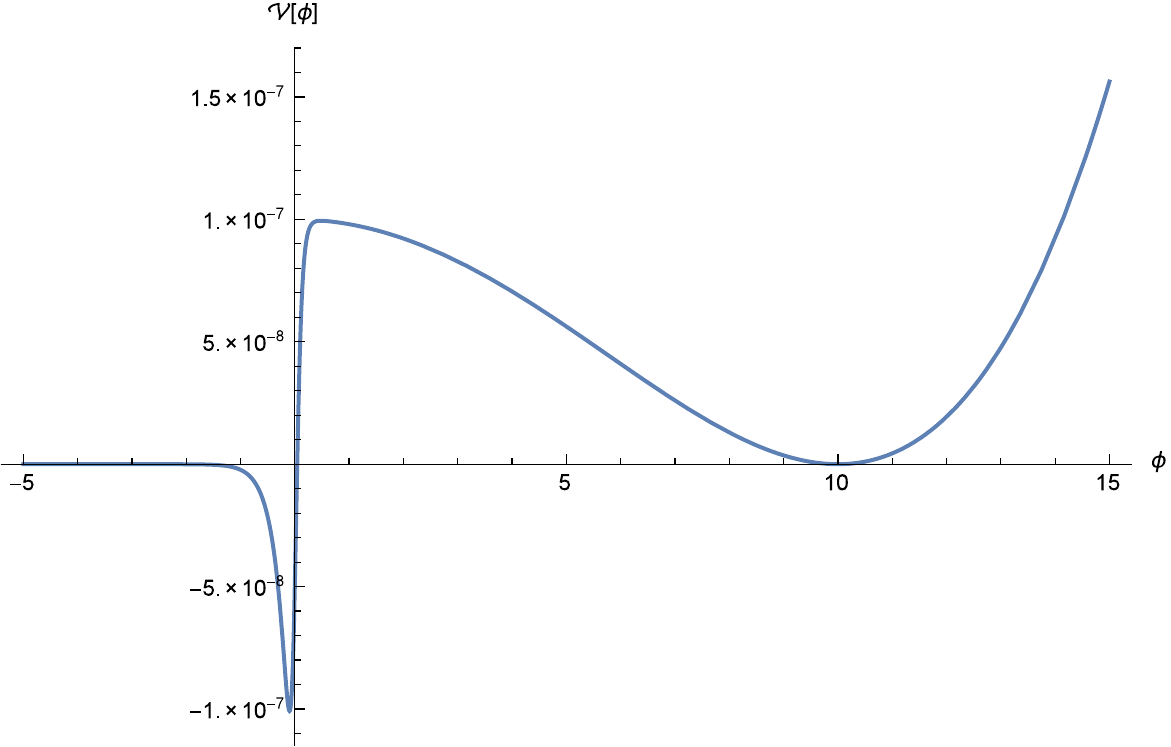}
\includegraphics[width=7cm,height=6cm]{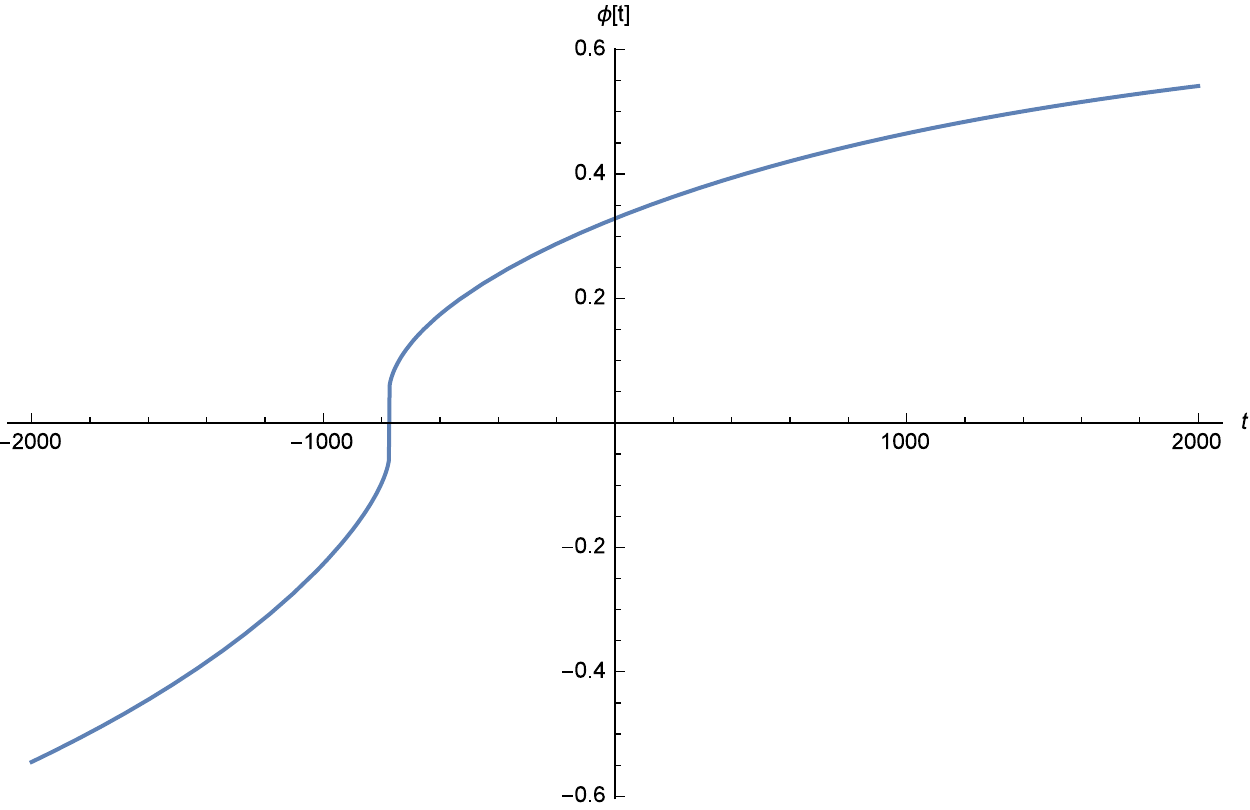}
\caption{Plots of functions ${\cal V}(\phi)$ and $\phi(t)$.}\label{Vphi}
\end{center}
\end{figure}

As has explained in the introduction (and see also~\cite{Qiu:2015nha}), by making use of Galileon form, this model has no ghost problem. Moreover, the anisotropy problem is also gone because of large $w_c$, while the scale-invariant power spectrum could be obtained at inflationary stage. Although recently people find that the such kind of Galileon bounce models has the problem of gradient instability~\cite{Libanov:2016kfc}, this usually happens round bounce region which, as has been shown before, will not have much effects to the whole phenomenological picture. Moreover, this instability can be cured by some operators which can come from the Effective Field Theory~\cite{Cai:2016thi, Cai:2017tku, Cai:2017dxl} (see also discussions in~\cite{Qiu:2015nha}). Models which contains these operators in a covariant way is presented in the following works of~\cite{Cai:2017dyi, Cai:2017pga}.

\section{Tensor perturbations}\label{sec_tensor}
Primordial gravitational waves are gravitational perturbations that are produced by the vacuum fluctuations of gravity. The linear tensor perturbations can be written as $g_{\mu\nu}=a^2(\tau)(\eta_{\mu\nu}+h_{\mu\nu})$. As a theoretical study, we briefly discuss the tensor perturbations generated in our model. The action of the perturbations can be expressed as
\begin{equation}\label{tensor}
S^2_T=\frac{1}{2}\sum_{s=+,\times}\int d\eta d^3xa^2\left[h_s^{\prime 2}-(\partial h_s)^2\right]~,
\end{equation}
where $h_s=h_+$ or $h_\times$, representing two independent states. It is convenient to calculate the perturbations with rescaled variables $v\equiv ah_{+,\times}/2$. From Eq.~(\ref{tensor}), and working in the momentum space, we can obtain
\begin{equation}\label{tensor_per}
v''_k+(k^2-\frac{a''}{a})v_k=0~.
\end{equation}

We follow the same method as what we did in \ref{scalar}. For the initial condition of $h$, we also choose the Bunch-Davies vacuum, so the solution of the tensor perturbation at the initial time is:
\begin{equation}
  v_k\sim \frac{1}{\sqrt{2k}}e^{-ik\eta}~.
\end{equation}
Substituting Eq.~(\ref{a3phase}) into Eq.~(\ref{tensor_per}), we can get the solution at constraint phase:
\begin{equation}\label{solution_tensor}
 v_k=\sqrt{-(\eta-\tilde{\eta}_{B-})}\{c_1^TH^{(1)}_{\nu_-}[-k(\eta-\tilde{\eta}_{B-})]+c_2^TH^{(2)} _{\nu_-}[-k(\eta-\tilde{\eta}_{B-})]\}~,~~~\nu_-\equiv\frac{(\epsilon_c-3)}{2(\epsilon_c-1)}~,
\end{equation}
and considering the initial conditions, we can get the coefficients:
\begin{equation}
  c_1^T=\frac{\sqrt{\pi}}{2}e^{i\frac{\pi}{2}(\nu_-+\frac{1}{2})}~,~c_2^T=0~,
\end{equation}
Unlike the scalar perturbations, the so called $Q$ and $c_s^2$ functions will be trivial in the case of tensor perturbations, even in the bouncing phase. So we can get the equation of $v_k$:
\begin{equation}
  v''_k+(k^2-\alpha a_B^2)=0
\end{equation}
and the solution is:
\begin{equation}\begin{split}
&v_k=c_3^Ta_B^{-1}\cos[l'(\eta-\eta_b)]+c_4^Ta_B^{-1}\sin[l'(\eta-\eta_B)]~,~~~ k>\sqrt{\alpha}a_B\\
&v_k=c_3^Te^{l'(\eta-\eta_B)}+c_4^Te^{-l'(\eta-\eta_B)}~,~~~k<\sqrt{\alpha}a_B
\end{split}\end{equation}
where $l'^2\equiv|k^2-\alpha a_B^2|$.
In the inflationary expanding phase, the solution of Eq.~(\ref{tensor_per})
\begin{equation}\label{tensor_inf}
v_k=\sqrt{-(\eta-\tilde{\eta}_{B+})}\{c_5^T H^{(1)}_{\nu_+}[-k(\eta-\tilde{\eta}_{B+})]+c_6^TH^{(2)}_{\nu_+}[-k(\eta-\tilde{\eta}_{B+})]\}~,~~~\nu_+\equiv\frac{(\epsilon_e-3)}{2(\epsilon_e-1)}~.
\end{equation}
According to the continuity of $v_k$, we can get the values of $c_3^T$, $c_4^T$, $c_5^T$ and $c_6^T$.
\begin{equation}\begin{split}
\label{tensor_c3c4}
c_3^T=&\frac{1+i}{4l'}\sqrt{\pi}\{\sqrt{-\frac{1}{\mathcal{H}_{\rm{con}}}}kH_1^{(1)}[-\frac{k}{2\mathcal{H}_{\rm{con}}}]\sin[l'(\eta_B-\eta_{B-})]\\
&+H_0^{(1)}[-\frac{k}{2\mathcal{H}_{\rm{con}}}]\{\sqrt{-\frac{1}{\mathcal{H}_{\rm{con}}}}l'\cos[l'(\eta_B-\eta_{B-})]-\sqrt{-\mathcal{H}_{\rm{con}}}\sin[l'(\eta_B-\eta_{B-})]\}\}~,\\
c_4^T=&\frac{1+i}{4l'}\sqrt{\pi}\{-\sqrt{-\frac{1}{\mathcal{H}_{\rm{con}}}}kH_1^{(1)}[-\frac{k}{2\mathcal{H}_{\rm{con}}}]\cos[l'(\eta_B-\eta_{B-})]\\
&+H_0^{(1)}[-\frac{k}{2\mathcal{H}_{\rm{con}}}]\{\sqrt{-\mathcal{H}_{\rm{con}}}\cos[l'(\eta_B-\eta_{B-})]+\sqrt{-\frac{1}{\mathcal{H}_{\rm{con}}}}l'\sin[l'(\eta_B-\eta_{B-})]\}\}~.\\
\end{split}\end{equation}
\begin{equation}\begin{split}
\label{tensor_c5c6}
c_5^T=&\frac{1}{8k^2}e^{i\frac{\pi}{4}-\frac{ik}{{\mathcal H}_{\rm{exp}}}}\pi\sqrt{-\frac{k}{{\mathcal H}_{\rm{con}}}}\\
&\times\frac{1}{l'}({\mathcal H}_{\rm{exp}}^2-i{\mathcal H}_{\rm{exp}}k-k^2)\{kH_1^{(1)}[-\frac{k}{2{\mathcal H}_{\rm{con}}}]+H_0^{(1)}[-\frac{k}{2{\mathcal H}_{\rm{con}}}][l'\cos(l'\Delta\eta_B)+{\mathcal H}_{\rm{con}}\sin(l'\Delta\eta_B)]\}\\
&+({\mathcal H}_{\rm{exp}}+ik)\{-kH_1^{(1)}[-\frac{k}{2{\mathcal H}_{\rm{con}}}]\cos(\Delta\eta_B)+H_0^{(1)}[l'\sin(l'\Delta\eta_B)-{\mathcal H}_{\rm{con}}\cos(l'\Delta\eta_B)]\}~,\\
c_6^T=&\frac{1}{\sqrt{2}(\frac{k}{{\mathcal H}_{\rm{exp}}})^{3/2}}(\frac{1+i}{8})e^{\frac{ik}{{\mathcal H}_{\rm{exp}}}}({\frac{k}{{\mathcal H}_{\rm{exp}}}})^{3/2}\pi\\
&\times\frac{1}{l'}({\mathcal H}_{\rm{exp}}^2-i{\mathcal H}_{\rm{exp}}k-k^2)\{\sqrt{-\frac{1}{{\mathcal H}_{\rm{con}}}}k\sin(l'\Delta\eta_B)H_1^{(1)}[-\frac{k}{2{\mathcal H}_{\rm{con}}}]\\
&+H_0^{(1)}[-\frac{k}{2{\mathcal H}_{\rm{con}}}][\sqrt{-\frac{1}{{\mathcal H}_{\rm{con}}}}l'\cos(l'\Delta\eta_B)-\sqrt{-{\mathcal H}_{\rm{con}}}\sin(l'\Delta\eta_B)]\}\\
&+({\mathcal H}_{\rm{con}}-ik)\{-\sqrt{-\frac{1}{{\mathcal H}_{\rm{con}}}}k\cos(l'\Delta\eta_B)H_1^{(1)}[-\frac{k}{2{\mathcal H}_{\rm{con}}}]\\
&+H_0^{(1)}[-\frac{k}{2{\mathcal H}_{\rm{con}}}][\sqrt{-{\mathcal H}_{\rm{con}}}\cos(l'\Delta\eta_B)+\sqrt{-\frac{1}{{\mathcal H}_{\rm{con}}}}l'\sin(l'\Delta\eta_B)]\}~.
\end{split}\end{equation}

The solution Eq.~(\ref{tensor_inf}) in the inflationary phase corresponds to the power spectrum that we could observe, and from the Eq.~(\ref{tensor_c3c4}) and (\ref{tensor_c5c6}), the power spectrum of curvature perturbation can be written as:
\begin{equation}
\label{spectrumT}
P_T\equiv2\frac{k^3}{2\pi^2}|h_k|^2=\Delta^2_{+,\times}|c^T_5-c^T_6|^2
\end{equation}
where
\be
\Delta^2_{+,\times}=\frac{2{\cal H}^2}{\pi^2M_p^2\epsilon_e}~,~~~
|c^T_5-c^T_6|^2\sim\left\{
\begin{array}{l}
k^{\frac{2\epsilon_c}{\epsilon_c-1}}~,~~~\text{for small $k$}~,\\\\
1+\text{trigonometric functions}~,~~~\text{for large $k$}~.
\end{array}\right.
\ee

According to the result for scalar and tensor perturbations, we can calculate the tensor-scalar-radio $r$, conventionally defined as $r\equiv P_T/P_{\mathcal{R}}$. From Eqs. (\ref{spectrum3}) as well as (\ref{spectrumT}), one gets
\be
\label{TSratio}
r=\frac{\Delta_{+,\times}^2}{\Delta_R^2}\frac{|c^T_5-c^T_6|^2}{|c_5-c_6|^2}=16\epsilon_e\times{\cal O}(1)~,
\ee
where in the last step, we made use of the fact that the modulation in scalar and tensor spectrum behaves in the same way. Therefore the tensor/scalar ratio is roughly the same as that in slow-roll inflation. For the recent constraints on the tensor/scalar ratio from the Planck data is about $r<0.07$ at $2\sigma$ C. L. \cite{Ade:2015tva}, the $r$ in Eq. (\ref{TSratio}) is acceptable provided $\epsilon_e\lesssim{\cal O}(10^{-3})$. As a side remark, this can also be applicable to the large scale region, although there is a blue tilt in both tensor and scalar power spectrum.

\section{Non-Gaussianites}\label{sec_non-gaussian}
In the above subsection, we studied the linear perturbation theory of the universe. However, non-Gaussianities also plays a crucial role in cosmological perturbations. Recent observations provided us a precise measurement of primordial non-Gaussianities, which implies a tight constraint of $f^{local}_{NL}=0.8\pm5.0$, $f^{equil}_{NL}=-4\pm43$, $f^{ortho}_{NL}=-26\pm21$ (combined temperature and polarization data, $68\%$ CL, statistical) \cite{Ade:2015ava}. Therefore, non-Gaussianities can also be treated as a powerful criteria to justify the early universe models. In this section, we discuss about the (equilateral) non-Gaussianities generated in the bounce inflation scenario.


In order to express the action Eq.~(\ref{lagrangian}) up to third order, we need to introduce an auxiliary filed $\xi$ to eliminate the perturbation parameter $\varphi$ in Eq.~(\ref{metric}). Follow the same lines of Ref.~\cite{DeFelice:2011uc, DeFelice:2011zh}, the $\xi$ satisfies:
\begin{equation}\label{def_phi}
  \varphi=-\frac{M_p^2}{M_p^2H-\dot{\phi}XG_X}\zeta+\frac{a^2}{M_p^2}\xi~,~~~\text{where}~~~\partial^2\xi=Q\dot{\zeta}
\end{equation}
where the value of $Q$ is different in the different evolution periods.

And now, we can take a approximate analysis for the sake of simplicity. When the universe is far away from the bounce point, i.e., $|\phi/M_p|\gg1$, we can get the Lagrangian as Eq.~(\ref{limit}). In this case, we have $Q\simeq2M_p^2\epsilon$ ($\epsilon=\epsilon_c$ for contracting phase, $\epsilon=\epsilon_e$ for inflation phase) and $c_s^2\simeq1$. The cubic action can be written:

\begin{equation}\begin{split}\label{action_third}
  S^{(3)}=&\int dtd^3x\{a^3\mathcal{Z}_1M_p^2\zeta\dot{\zeta}^2+a\mathcal{Z}_2M_p^2\zeta(\partial{\zeta})^2+a^3\mathcal{Z}_3\dot{\zeta}(\partial_i\zeta)(\partial_i\xi)+a^3(\mathcal{Z}_4/M_p^2)\partial^2\zeta(\partial{\xi})^2\\
  &-2[\partial_k\zeta\partial_k\xi-\partial^{-2}\partial_i\partial_j(\partial_i\zeta\partial_j\xi)-\dot{\zeta}\zeta-\frac{(\partial\zeta)^2-\partial^{-2}\partial_i\partial_j(\partial_i\zeta\partial_j\zeta)}{4a^2}][2\frac{d}{dt}(a^3M_p^2\epsilon\dot{\zeta})-a  \partial^2\zeta]\}
\end{split}\end{equation}
where the coefficients $\mathcal{Z}_i(i=1,2,3,4)$ are
\begin{equation}\begin{split}
  \mathcal{Z}_1=&\epsilon-\frac{\dot{\epsilon}}{H\epsilon}\\
  \mathcal{Z}_2=&\epsilon+\frac{\dot{\epsilon}}{H\epsilon}-2\frac{\dot{c_s^2}}{H}\\
  \mathcal{Z}_3=&\frac{\epsilon^2}{2}-2\epsilon\\
  \mathcal{Z}_4=&\frac{1}{4}\epsilon
\end{split}\end{equation}
The last term of Eq.~(\ref{action_third}) survives only at second order in $\zeta$, and we will neglect its contribution to non-Gaussian. The $\zeta(k)$ is primordial quantum perturbation in the early universe, we can obtain the vacuum expectation value of $\zeta(k)$ for three-point operator by using the interaction picture.
\begin{equation}\label{interacting}
  <\zeta(k_1)\zeta(k_2)\zeta(k_3)>=-i\int_{\eta_i}^{\eta_e}d\eta a<0|[\zeta(\eta_e,k_1)\zeta(\eta_e,k_2)\zeta(\eta_e,k_3),\mathcal{H}_{int}(\eta)]|0>
\end{equation}
where $\mathcal{H}_{int}(\eta)$ is the interaction Hamiltonian which is equal to the Lagrangian of the cubic action \cite{Maldacena:2002vr, Chen:2006nt}.  After going over to Fourier space, one finds that
\begin{equation}\label{Fourier}
  <\zeta(k_1)\zeta(k_2)\zeta(k_3)>=(2\pi)^3\delta^3(k_1+k_2+k_3)(P_\mathcal{R})^2F_\zeta(k_1,k_2,k_3)
\end{equation}
where $P_\mathcal{R}$ is the power spectrum of perturbation in the contracting or inflationary phase and the parameter $F_\zeta$ can be defined as
\begin{equation}
  F_\zeta(k_1,k_2,k_3)=\frac{(2\pi)^4}{\Pi_{i=1}^3}\mathcal{A}_\zeta(k_1,k_2,k_3)
\end{equation}
where
\begin{equation}\label{A_r}
  \mathcal{A}_\zeta=\frac{1}{4\epsilon}(\mathcal{S}_1\mathcal{Z}_1+\mathcal{S}_2\mathcal{Z}_2)+\frac{1}{8}\mathcal{S}_3\mathcal{Z}_3+\frac{\epsilon}{4}\mathcal{S}_4\mathcal{Z}_4
\end{equation}
in which $\mathcal{S}_1,\mathcal{S}_2,\mathcal{S}_3$ and $\mathcal{S}_4$ are the shape functions with the relations
\begin{equation}\begin{split}
  &\mathcal{S}_1=\frac{2}{K}\sum_{i>j}k_i^2k_j^2-\frac{1}{K}\sum_{i\neq j}k_i^2k_j^2, \hspace{3mm} \mathcal{S}_2=\frac{1}{2}\sum_ik_i^3+\frac{2}{K}\sum_{i>j}k_i^2k_j^2-\frac{1}{K^2}\sum_{i\neq j}k_i^2k_j^3\\
  &\mathcal{S}_3=\sum_ik_i^3-\frac{1}{2}\sum_{i\neq j}k_ik_j^2-\frac{2}{K^2}\sum_{i\neq j}k_i^2k_j^3,\hspace{3mm}\mathcal{S}_4=\frac{1}{K^2}[\sum_ik_i^5+\frac{1}{2}\sum_{i\neq j}k_ik_j^4-\frac{3}{2}\sum_{i\neq j}k_2^2k_j^2-k_1k_2k_3\sum_{i>j}k_ik_j]
\end{split}\end{equation}
where $K = k_1+k_2+k_3$.

In general, the non-linear parameter $f_{NL}$, characterizing the amplitude of non-Gaussian, can be defined as \cite{Spergel:2003cb, Komatsu:2008hk}
\begin{equation}
  f_{NL}=\frac{10}{3}\frac{\mathcal{A}_\zeta}{\sum_{i=1}^{3}k_i^3}~.
\end{equation}

As the model containing higher derivative term, the equilateral shape of the non-Gaussianity where $k_1=k_2=k_3=k$ will be the most significant. The non-Gaussian parameter of equilateral bispectrum Eq.~(\ref{A_r})
\begin{equation}\begin{split}
  &f_{NL}^{equil}=\frac{55}{36}\epsilon_e+\frac{5\dot{\epsilon_e}}{12H\epsilon_e}~~~\text{for large $k$}~.
\end{split}\end{equation}
As one can see that, since we assume that we can only observe the perturbations of large $k$ modes,  $f^{equil}_{NL}$ will be suppressed according to $\epsilon_e$. For $\epsilon_e\lesssim {\cal O}(0.01)$, the non-Gaussianity is quite within the observational constraints.




\end{document}